\documentclass[aps,pre,floats,twocolumn,showpacs,superscriptaddress]{revtex4}
\usepackage{amssymb}
\usepackage{graphicx}% Include figure files
\usepackage{amsmath}
\usepackage{enumerate}
\usepackage[usenames,dvipsnames]{color}

\begin{document}

%\begin{frontmatter}

\title{Stability and robustness analysis of cooperation cycles driven by
destructive agents in finite populations}

\author{Rub\'en J. Requejo}
\affiliation{Department de F\'{\i}sica,
Universitat Aut\'{o}noma de Barcelona, 08193 Bellaterra, Barcelona, Spain}

\author{Juan Camacho}
\affiliation{Department de F\'{\i}sica,
Universitat Aut\'{o}noma de Barcelona, 08193 Bellaterra, Barcelona, Spain}

\author{Jos\'e A. Cuesta}
\affiliation{Grupo Interdiciplinar de Sistemas Complejos
(GISC),Departamento de Matem\'{a}ticas,
Universidad Carlos III de Madrid, 28911 Legan\'{e}s, Madrid, Spain}

\affiliation{Instituto de Biocomputaci\'on y F\'\i sica de Sistemas Complejos
(BIFI), Universidad de Zaragoza, Campus R\'\i o Ebro, 50018 Zaragoza, Spain}

 \author{Alex Arenas}
\affiliation{Departament d'Enginyeria Inform{\`a}tica i Matem{\`a}tiques,
Universitat Rovira i Virgili, 43007 Tarragona,
  Spain}

\begin{abstract}
The emergence and promotion of cooperation is one of the main issues in
evolutionary game theory, as cooperation is amenable to exploitation by
defectors, which take advantage from cooperative individuals at no cost
dooming them to extinction. It has been recently shown that the existence
of purely destructive agents (termed jokers) acting
on the common enterprises (public goods games), can induce stable
limit cycles between cooperation, defection and destruction when infinite
populations are considered. These cycles allow for time lapses in which
cooperators represent a relevant fraction of the population, providing
a mechanism for the emergence of cooperative states in nature and human
societies. Here we study analytically and through agent-based simulations
the dynamics generated by jokers in finite populations for several
selection rules. Cycles appear in all cases studied thus showing
that the joker dynamics generically yields a robust cyclic behavior 
not restricted to infinite populations.
We have also computed the average time in which the population consists
mostly of just one strategy and compare the results with numerical
simulations.

\end{abstract}


\maketitle

\section{Introduction}

Cooperation is necessary for the appearance of complex structures and
higher order selection units from individual behaviors. In this way,
cooperation between unicellular life forms gave rise to multicellular organisms,
cooperative animals form communities and cooperation between humans
gives rise to the complex societies we live in \citep{maynard-smith:1995}. Thus,
it is very important to understand the conditions that allow cooperation
to thrive and evolve in nature and society.
However, cooperative behaviors are not stable, as they are easily
invaded by selfish individuals, who benefit from the interactions with
cooperative ones but avoid paying the costs attached to cooperation. The
selfish individuals, called defectors, have a higher fitness ---a
measure of reproductive success--- than cooperators in any well mixed
population, and therefore will spread under the action of natural
selection, leading to the extinction of the cooperative behavior
\citep{hofbauer:1998}, and to populations where nobody benefits from
altruistic acts (the ``tragedy of the commons'' \citep{hardin:1968}).

In the last decades the study of the Public Goods (PG) game, a
mathematical metaphor of a common enterprise, in which cooperative
individuals invest ---pay a cost--- and share the benefits with
all the players involved in the game, has led to the discovery of
some mechanisms that allow cooperation to thrive, as introducing
reputation \citep{milinski:2005}, diversity in number and size
of groups \citep{santos:2008}, linking group size and payoffs
\citep{wakano:2009}, or the inclusion of spatial structure and conditional
behaviors \citep{roca:2009c,szolnoki:2011,szolnoki:2012}. Furthermore, it has
been proven that the introduction of some behavioral types in well-mixed
population, as punishers
\citep{fehr:2002, hauert:2007} or individuals which do not participate
in the PG and instead receive a fixed benefit (the so-called `loners')
\citep{hauert:2002b}, may promote
cooperation. However, the only behavioral type found so far that allows
for the emergence of stable cycles in the presence of mutations is the
so called joker strategy \citep{arenas:2011}. Jokers do not take part
of the benefits produced by the PG game and instead they provoke a
damage to the common enterprise, thus affecting every individual involved
in it. Surprisingly, the effect of these indiscriminate destructive
agents on the dynamics is the induction of robust evolutionary stable
limit cycles of cooperation, defection and destruction.
A cyclic dynamics can also be found when loners are involved. However,
jokers induce cycles that are dynamically different from those found with
loners \citep{hauert:2002b}, because the latter are neutrally stable
(i.e, have no fixed amplitudes) and disappear in the presence of mutations
or structural noise.

The inclusion of destructive agents in the PG game is motivated by
the observation that, in nature and society, the appearance of a risk, as
those created by common enemies, predators or simply dangerous situations,
may induce cooperation among the victims \citep{lavalli:2009, krams:2010}.

In a previous
work \citep{arenas:2011}, the stability of the evolutionary cycles induced
by jokers was proven for infinite populations through
the analysis of the replicator mutator equation. Here, we extend the study
to finite populations and analyze the dynamics for different updating
methods \citep{szabo:2007}, i.e., different selection dynamics for the
population, in order to analyze if robust cycles are also found. The
conclusion we reach from this study is that
the robust cycles obtained using the replicator-mutator equation
are not just restricted to this particular dynamics but
are a generic feature of the joker model.

The paper is organized as
follows. In section II we explain the PG game and review the dynamics
in infinite populations. Section III provides the stochastic equations 
describing the evolutionary dynamics for finite populations. 
In section IV we analyze the joker dynamics in
finite populations using different selection dynamics in order to check
the existence of cycles. Section V is devoted to conclusions.

\begin{figure*}
\begin{center}
\includegraphics[width=170mm,clip=]{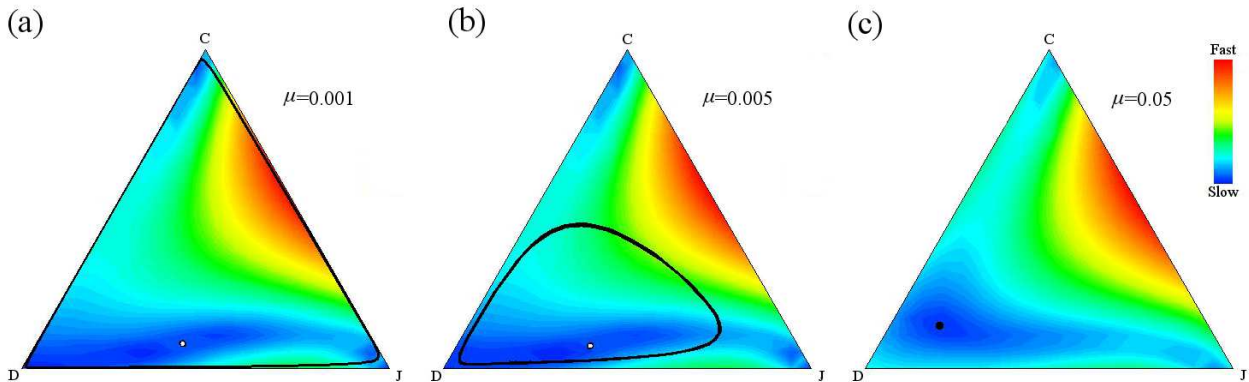}
\end{center}
\caption{(Color online.) Cycles induced by jokers in infinite populations.
The simplexes 
describe the replicator-mutator dynamics for a population of cooperators,
defectors and jokers with parameter values satisfying $n>r>1+d(n-1)$,
for which a rock-paper-scissor dynamics is expected. For small mutation rates, the only equilibrium is a repeller (white dot in (a),(b)), and
trajectories end up in a stable limit cycle of decreasing amplitude with
increasing $\mu$ (black line); when mutations reach a critical value $\mu_c$,
the system undergoes a Hopf bifurcation and a stable mixed equilibrium appears
(black dot in (c)).
Thus the presence of jokers induces periodic bursts of cooperation for low
mutation rates, and stable coexistence for high $\mu$.
Parameters: $n=5$, $r=3$, $d=0.4$, $\mu$ is (a) 0.001, (b) 0.005, (c) 0.05.
(Images generated using a modified version of the Dynamo Package
\citep{sandholm:2007}).}
\label{fig:simplex}
\end{figure*}

\section{Evolutionary cycles induced by jokers in Public Good games}

In each PG game a number $n$ of individuals is randomly chosen from
the entire population. Each cooperative individual contributes to the
common enterprise at a cost $c$ to itself, which yields a benefit $b=rc$
($r>1$) equally distributed between all players; defectors free-ride the
public good at no cost, thus obtaining a higher benefit than cooperators;
jokers do not participate on the benefits, and each one provokes a damage
$-d<0$ to be shared by all individuals engaged in the PG game. Note that the
cost paid by C players can be set to $c=1$ without loss of generality:
all other payoffs are given in units of $c$.

Let us call $0\le m \le n$ the number of cooperative participants,
$0\le j\le n$ the number of jokers, $n-m-j\ge 0$
the number of defectors and $S=n-j$ the number of non-jokers, i.e.,
the number of individuals involved in the PG game,
and that potentially benefit from the PG.
Then the payoff of a defector will be
$\Pi_{\rm D}(m,j)= (rm-dj)/S$, and that of a cooperator
$\Pi_{\rm C} =\Pi_{\rm D}-1$; as previously stated, in each interacting
group
defectors will always do better than cooperators. Jokers' payoff is
always $0$.

In summary, we have
\begin{align}
&\Pi_{\rm D}(m,j) =\frac{rm-dj}{S}, && 0\le m\le S-1\le n-1, \nonumber \\
&\Pi_{\rm C}(m,j) =\frac{rm-dj}{S}-1, && 1\le m\le S\le n, \label{eq:payoffs} \\
&\Pi_{\rm J}(m,j) =0, && 0\le S\le n-1. \nonumber
\end{align}

A simple invasion analysis \cite{arenas:2011} shows that the PG thus
defined determines a tragedy of the commons \cite{hardin:1968} for
$r<r_{\rm max}=n(M-1)/(M-n)$, where $M$ is the population size. For infinite
populations, the latter condition reduces to $r<n$, as usual in PG games.
Therefore, a mixed population of cooperators
and defectors with $r<r_{\rm max}$
will end up composed of defectors only. The interesting question is then
if jokers may prevent the
extinction of cooperators and under which conditions. In
Ref.~\cite{arenas:2011}, it was shown
that, in the region of interest, namely $1<r<r_{\rm max}$, $d>0$, the
system exhibits:
(a) joker-cooperator bistability for $1+d/(M-1)<r<1+(n-1)d$,
(b) joker dominance for $r<1+d/(M-1)$ and, most importantly, (c) a
rock-paper-scissor (RPS) cyclic dominance of the three strategies for
\begin{equation}
\label{ineq}
r>1+(n-1)d.
\end{equation}
This condition expresses the fact that a single cooperator gets
a positive payoff in spite of the damage inflicted by $n-1$ jokers, which
allows cooperators to thrive in the damaging environment that represents
a population of jokers and re-establish a cooperative environment. In
Ref.~\cite{arenas:2011},
we analyzed just one dynamics, namely the replicator-mutation
dynamics, and showed that it produces stable (robust)
limit cycles C$\to$D$\to$J$\to$C when mutations are rare, and
stable coexistence for high mutation rates (Fig.~\ref{fig:simplex}). In the
following we will analyze 
the dynamics of PG with jokers in finite populations under different update
rules, check the appearance of cycles and calculate the average time spent in
each homogeneous state, in order to decide which dynamics better promotes the
survival of cooperation.

\section{Stochastic dynamics in finite populations}
\label{sec:finite}

The deterministic evolution represented by the replicator-mutator equation
is an idealization of the system behavior in the limit of infinite populations.
To get a deeper insight into the model we need to address the question
what happens when populations have a finite size $M$. To begin with we need
to describe the microscopic dynamics in more detail. Hauert et
al.~\cite{hauert:2007} have proposed a protocol in which random selections
of $n$ players are gathered together to play the game. After receiving their
corresponding payoffs the group dissolves and a new one is sampled. This
sampling is made a sufficient number of times so that on average each player
receives a payoff proportional to the mean payoff she can obtain given
the composition of the population.

Suppose there are $m$ cooperators, $j$ jokers, and $M-m-j$ defectors in
the population. The probability that the sampling of $n$ individuals contains
$k$ cooperators, $l$ jokers, and $n-k-l$ defectors is given by the
extended hypergeometric distribution
\begin{equation}
p(k,l|n,m,j,M)=\frac{\displaystyle
\binom{m}{k}\binom{j}{l}\binom{M-m-j}{n-k-l}}{\displaystyle
\binom{M}{n}}.
\label{eq:general-hyper}
\end{equation}
The average payoff of strategy X within this population, $P_{\rm X}(m,j)$,
is obtained by
averaging formulae \eqref{eq:payoffs} with this probability distribution.
This is done in Appendix~\ref{app:B}, where explicit expressions for
$P_{\rm C}(m,j)$ and $P_{\rm D}(m,j)$ are obtained ---obviously
$P_{\rm J}(m,j)=0$ irrespective of the population composition.

Once payoffs are obtained evolution proceeds by imitation. Different
payoff-dependent updating rules have been proposed in the literature \cite{szabo:2007}.
All of them describe a process of birth and death which is defined by the
transition probability $T(m',j'|m,j)$ from a population with composition
$(m,j)$ to another one with composition $(m',j')$ within the set
\begin{equation}
\begin{split}
\mathcal{N}_{m,j}=&\, \{(m,j),(m\pm 1,j),(m,j\pm 1), \\
&(m+1,j-1),(m-1,j+1)\}.
\end{split}
\end{equation}
If now $\Pi(m,j;t)$ denotes the probability that the population has a
composition given by $(m,j)$ at time $t$, then this probability evolves
according to
\begin{equation}
\Pi(m,j;t+1)=\sum_{(m',j')\in\mathcal{N}_{m,j}}T(m,j|m',j')\Pi(m',j';t).
\label{eq:iteration}
\end{equation}
It is implicitly asumed that $\Pi(m,j;t)=0$ for all $t$ if the pair $(m,j)$ is
outside the set $\mathcal{P}\equiv\{(m,j)\in\mathbb{Z}^2:m,j\ge 0$, $m+j\le M\}$.

If we introduce matrix $\mathbf{T}$, with elements $T(m,j;m',j')$ [$(m,j)$ is
the ``row index'' and $(m',j')$ the ``column index''] defined as
\begin{equation}
T(m,j;m',j')=
\begin{cases}
T(m,j|m',j') & \text{if $(m',j')\in\mathcal{N}_{m,j}$,} \\
0 & \text{otherwise,}
\end{cases}
\end{equation}
and vectors $\boldsymbol\Pi(t)$, with elements $\Pi(m,j;t)$ [where
$(m,j),(m',j')\in\mathcal{P}$], then Eq.~\eqref{eq:iteration} can be cast in
matrix notation simply as
\begin{equation}
\boldsymbol\Pi(t+1)=\mathbf{T}\,\boldsymbol\Pi(t).
\label{eq:PiTPi}
\end{equation}

\subsection{Stationary state}

If the process undergoes mutations then matrix $\mathbf{T}$ is ergodic
and Eq.~\eqref{eq:PiTPi} has got a unique stationary state, $\boldsymbol\pi$,
which is obtained by solving the linear system
\begin{equation}
\boldsymbol\pi=\mathbf{T}\boldsymbol\pi.
\label{eq:pistat}
\end{equation}
In the absence of mutations, though,
there are three absorbing states corresponding to the three homogeneous populations.
A homogeneous population remains invariant because the imitation process cannot
change its composition. We will denote these vectors $\mathbf{e}_{\rm C}$,
$\mathbf{e}_{\rm D}$, $\mathbf{e}_{\rm J}$, the index denoting the strategy
of the homogeneous population. Clearly
$e_{\rm C}(m,j)=\delta_{m,M}\delta_{j,0}$,
$e_{\rm D}(m,j)=\delta_{m,0}\delta_{j,0}$,
$e_{\rm J}(m,j)=\delta_{m,0}\delta_{j,M}$.

\subsection{Infinitely small mutation rate}

After every imitation attempt (whether successful or not), individuals can
randomly mutate their strategy. With probability $2\mu$ the actor of the
imitation event changes its current strategy into one of the other two
equally likely. Parameter $\mu$ is referred to as the mutation ratio. In this
section we will be concerned with mutation rates $\mu\ll 1$.

In the the limit $\mu\to 0^+$ the stationary probability distribution must
be a linear combination of the stationary vectors of the process without
mutations, so in principle, taking the limit
\begin{equation}
\lim_{\mu\to 0^+}\boldsymbol\pi=\sum_{{\rm X}={\rm C},{\rm D},{\rm J}}
\alpha_{\rm X}\,\mathbf{e}_{\rm X}
\label{eq:alpha}
\end{equation}
should provide the coefficients $\alpha_{\rm X}$ of this linear combination,
but this limit cannot be obtained directly from Eq.~\eqref{eq:pistat}. There is
an alternative though. It has been proven \cite{fudenberg:2006} that the
$\mu\to 0^+$ limit of this process is equivalent to another process with three
states, C, D, J, in which the transition probability between X and Y is equal
to the probability that a single mutant of type Y invades an otherwise
homogeneous population of X individuals, thus transforming it into a
homogeneous population of Y individuals. Intuitively, this is tantamount to
saying that mutations are so rare that the ultimate fate of a mutant is decided
before the next mutation occurs. The stationary vector in this space,
\begin{equation}
\boldsymbol\alpha=(\alpha_{\rm C},\alpha_{\rm D},\alpha_{\rm J}),
\end{equation}
provides the values of the coefficients $\alpha_{\rm X}$ in
\eqref{eq:alpha}.

Following \cite{hauert:2007}, let $\rho_{\rm YX}$ denote the probability
that a single Y mutant takes over the population made of the mutant and
$M-1$ individuals of type X. Then the transition probability of going
from state X to a different state Y in the three-states Markov chain
defined above will be $r_{\rm YX}=\rho_{\rm YX}\mu$. Introducing
$\mathbf{R}=(r_{\rm YX})$ so that the elements in each column add
up to one (this fixes the diagonal of the matrix), we can rewrite this
matrix as
$\mathbf{R}=\mathbf{I}+\mu\mathbf{Q}$, where
\begin{equation}
\mathbf{Q}=
\begin{pmatrix}
-\rho_{\rm DC}-\rho_{\rm JC} & \rho_{\rm CD} & \rho_{\rm CJ} \\
\rho_{\rm DC} & -\rho_{\rm CD}-\rho_{\rm JD} & \rho_{\rm DJ} \\
\rho_{\rm JC} & \rho_{\rm JD} & -\rho_{\rm CJ}-\rho_{\rm DJ}
\end{pmatrix}.
\end{equation}
Vector $\boldsymbol\alpha$ is then the solution of the linear
system $\mathbf{Q}\boldsymbol\alpha=\mathbf{0}$.
A little bit of algebra leads to the result
\begin{align}
\alpha_{\rm C} &= (\rho_{\rm CD}\rho_{\rm CJ}+\rho_{\rm CD}\rho_{\rm DJ}
                  +\rho_{\rm CJ}\rho_{\rm JD})/A, \\
\alpha_{\rm D} &= (\rho_{\rm DC}\rho_{\rm DJ}+\rho_{\rm DC}\rho_{\rm CJ}
                  +\rho_{\rm DJ}\rho_{\rm JC})/A, \\
\alpha_{\rm J} &= (\rho_{\rm JC}\rho_{\rm JD}+\rho_{\rm JC}\rho_{\rm CD}
                  +\rho_{\rm JD}\rho_{\rm DC})/A,
\end{align}
with $A$ chosen so as to fulfill 
\begin{equation}
\label{eq:normalpha}
\sum_{{\rm X}={\rm C},{\rm D},{\rm J}}
\alpha_{\rm X} = 1.
\end{equation}

\subsection{Finite mutation rates}

If the mutation rate is not zero the Markov chain is ergodic and
the stationary state can be obtained by solving numerically Eq.~\eqref{eq:pistat}.
This is accomplished with better accuracy by splitting
\begin{align}
\mathbf{T} &=\mathbf{T}_0+\mathbf{T}_1, \\
\boldsymbol\pi &=\sum_{{\rm X}={\rm C},{\rm D},{\rm J}}\alpha_{\rm X}\,
\mathbf{e}_{\rm X}+\boldsymbol\pi_1,
\label{eq:T1}
\end{align}
with $\mathbf{T}_0$ the transition matrix in the absence of mutations---i.e.,
with transitions describing only the imitation process. 
Then $\boldsymbol\pi_1$ is the solution of the linear system
\begin{equation}
(\mathbf{I}-\mathbf{T})\boldsymbol\pi_1=\sum_{{\rm X}={\rm C},{\rm D},{\rm J}}
\alpha_{\rm X}\,\mathbf{T}_1\mathbf{e}_{\rm X}.
\label{eq:pi1}
\end{equation}

\subsection{Imitation rules}
\label{sec:imitation}

In order to specify the transition matrix $\mathbf{T}$ we need to
describe the imitation process. Of the many different rules applied
in the literature \cite{szabo:2007} we have chosen the three most
commonly employed: unconditional
imitation, proportional update, and a Moran process.
In all cases the corresponding
matrix $\mathbf{T}$ is obtained in Appendix~\ref{app:C}.

Under unconditional imitation two players are chosen at random among
the population, one as the focal player and the other one as the model to
imitate. The focal player compares both payoffs and changes her strategy
to that of the model if the latter has a higher
payoff. In this case, the strategy with the highest fitness never changes
except by mutation, which is the only source of stochasticity in this rule.

Appendix~\ref{app:A} discusses the value $\boldsymbol{\alpha}$ for this
update rule. There are two possibilities:
\begin{enumerate}[(i)]
\item $r>1+(n-1)d$. In this cases all three homogeneous states are
equally likely [c.f.~Eq.~\eqref{eq:ui1}].
\item $r<1+(n-1)d$. In this cases J is the only absorbing state of
the process [c.f.~Eq.~\eqref{eq:ui2}].
\end{enumerate}

Proportional update is entirely similar to unconditional imitation
with the exception that imitation occurs with probability proportional
to the payoff difference between the model and the focal players. For
this reason the values of $\boldsymbol{\alpha}$ for this rule are the same as
those for unconditional imitation.

In a Moran process a strategy is chosen to be imitated (or reproduced)
with a probability proportional to its population-dependent fitness. The
player who imitates (or is replaced by the offspring of) this selected
player is randomly chosen from the rest of the population. The only
drawback of this rule is that fitnesses must be positive for it to
make sense, so they cannot be directly the payoffs of the game, because
they can take negative values. A standard mapping between payoff and
fitness is obtained by introducing the \emph{selection strength} $s$
\cite{nowak:2004a}.
This weights the contribution of the game to the total fitness of the
strategy as $F=1-s+sP$, with $P$ the average payoff. Bounding the value
of $s$ we can force $F$ to be positive.

The Moran process thus described defines a birth-death process with
two absorbing states, and the corresponding probabilities $\rho_{\rm XY}$
are obtained via standard formulae (see Appendix~\ref{app:A}).

\section{Results: Robustness of the cycles using different selection dynamics}
\label{results}

\begin{figure*}
\begin{tabular}{cc}
\includegraphics[width=88mm]{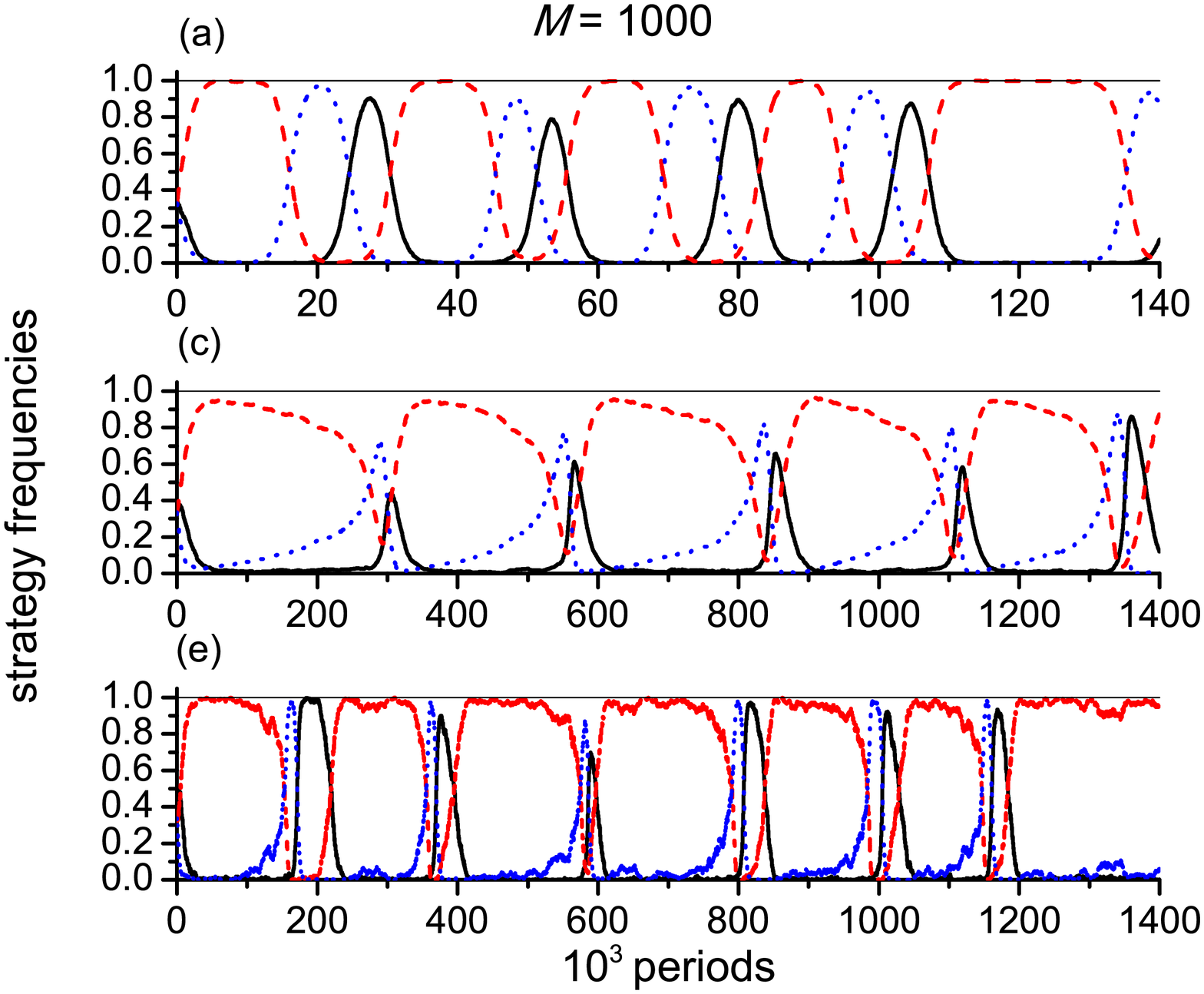} &
\includegraphics[width=88mm]{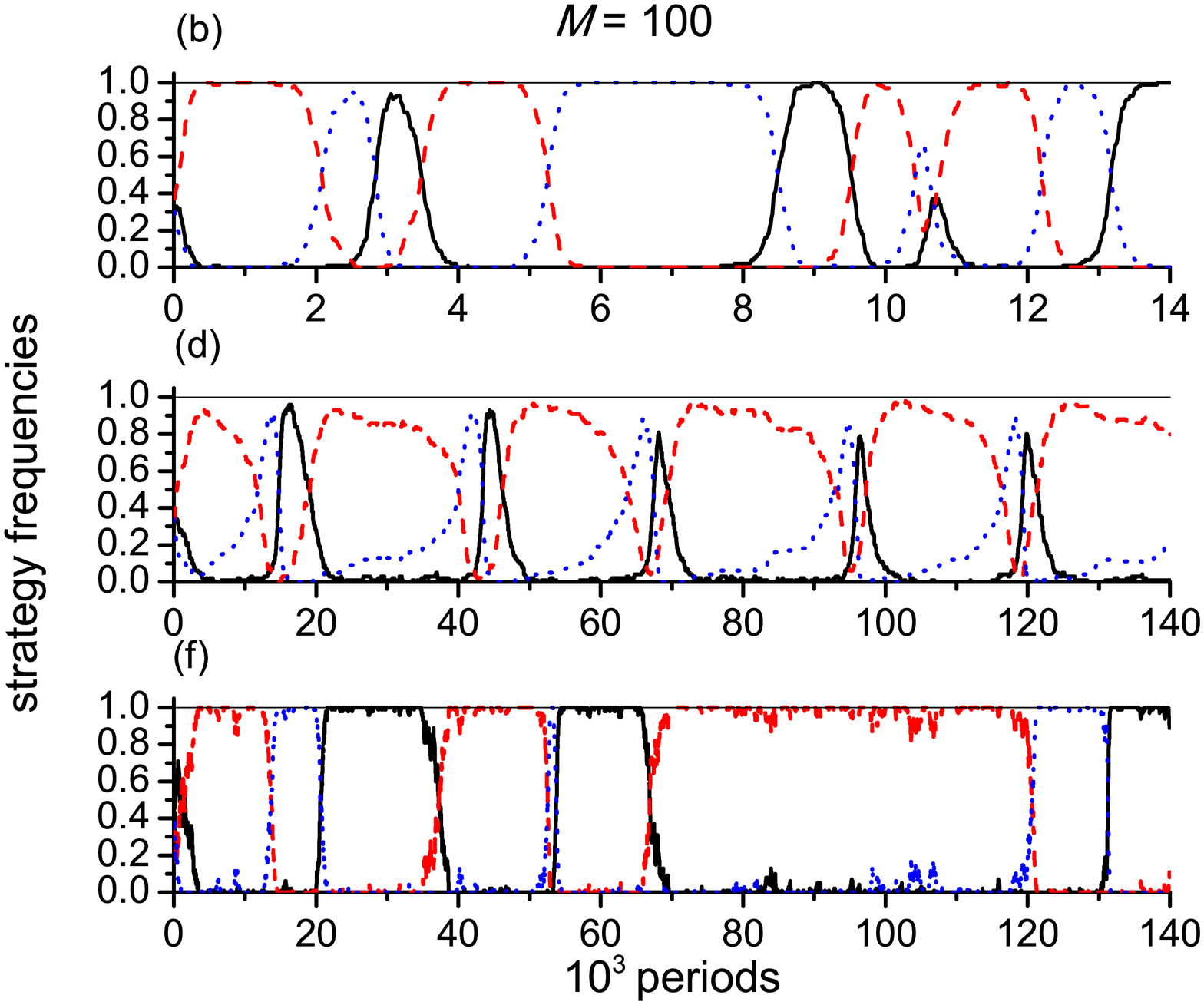}
\end{tabular}
\caption{(Color online.) Time evolution of the frequencies of the three
strategies in a population of $M=1000$ (left) and $M=100$ (right) individuals
playing a PG with jokers with different update rules: (a), (b) unconditional
imitation, (c), (d) proportional update, and (e), (f) a Moran process. The
presence of jokers induces a cyclic behavior irrespective of the update rule
and the population size, as long as the mutation rate $\mu>0$.
Black solid lines: cooperators, red dashed lines: defectors, and
blue dotted lines: jokers. One period corresponds to one updating event according to the evolutionary rule used. Parameters $n=5,r=3,d=0.4$,
$\mu=0.001$; in (a)-(d) $s=1$, in (e),(f) $s=0.38$.}
\label{fig:updatecomp}
\end{figure*}

In this section we compare the results of agent-based simulations
with those obtained by solving the stationary equation \eqref{eq:pistat}.
Simulations implement the following stochastic process. We
start with a population of $M$ individuals with equal amounts of C, D and
J players. Then:
\begin{enumerate}
\item Assuming that every time step each individual plays many
rounds of the game with different, randomly gathered groups of $n$ players,
the payoffs they obtain will be proportional to the average payoffs, as
calculated in Appendix \ref{app:B}. Thus we assume that these expressions
provide the payoffs each individual gains every time step.
\item These payoffs are used to update the population according to the
corresponding imitation rule. We implement the three rules described in
Sec.~\ref{sec:imitation}.
\item With probability $\mu$ each newborn mutates to a different strategy
(any of the other two with equal probability).
\end{enumerate}

A quite general result is that, irrespective of the population size,
at low mutation rates simulations show patterns of cyclic invasions
C$\to$D$\to$J$\to$C (see Fig.~\ref{fig:updatecomp}). These patterns
resemble the limit cycles observed in the replicator dynamics, i.e.,
for infinite populations \cite{arenas:2011} (c.f.~Fig.~\ref{fig:simplex}).

\begin{figure*}
\begin{tabular}[t]{cc}
(a) & (b)\\
\includegraphics[width=88mm]{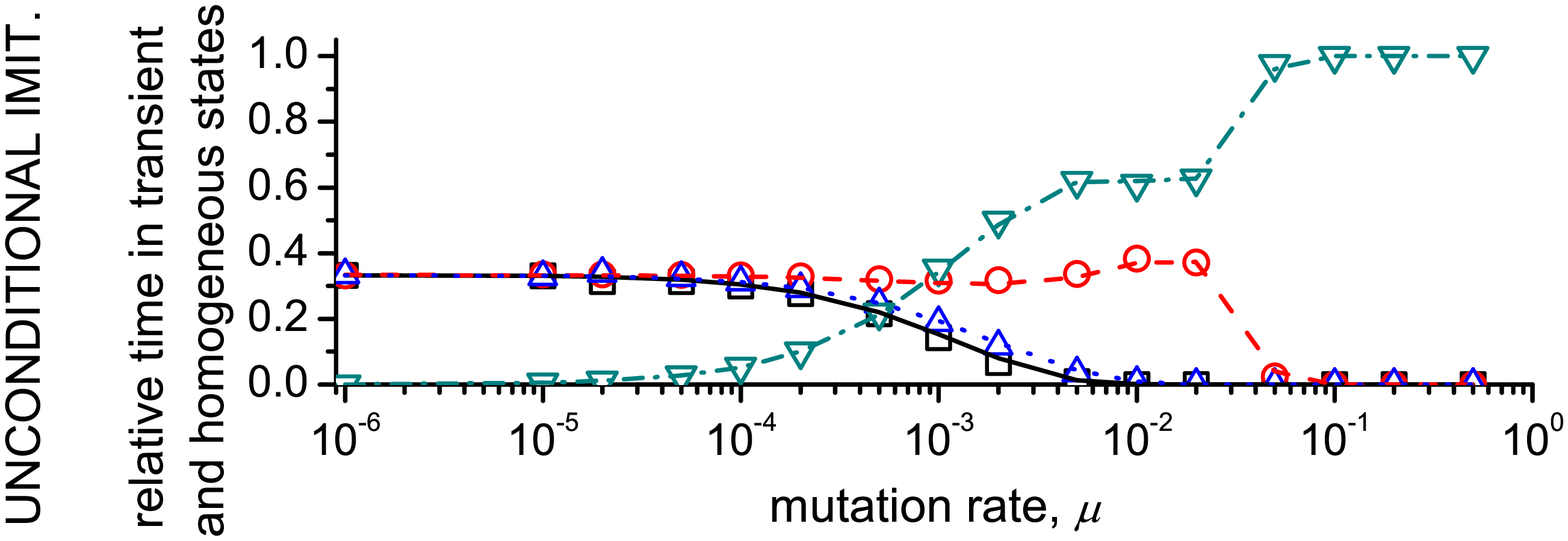} &
\includegraphics[width=88mm]{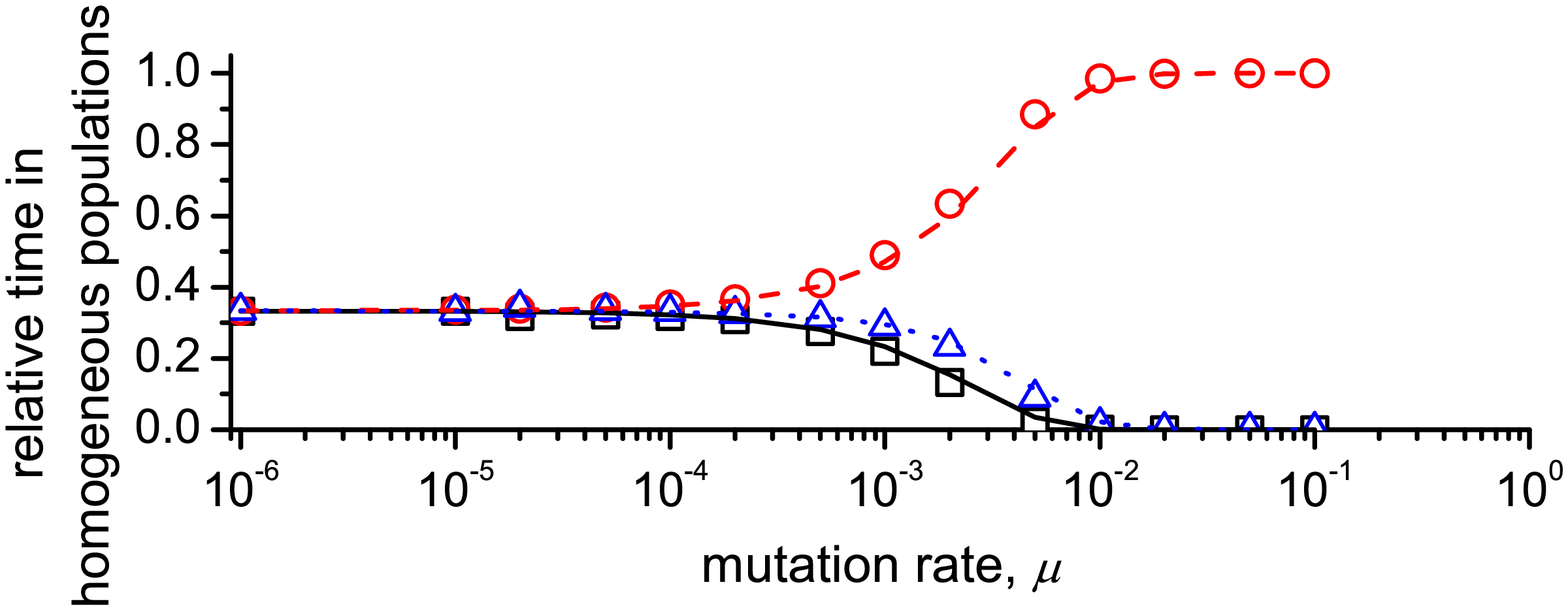} \\
(c) & (d) \\
\includegraphics[width=88mm]{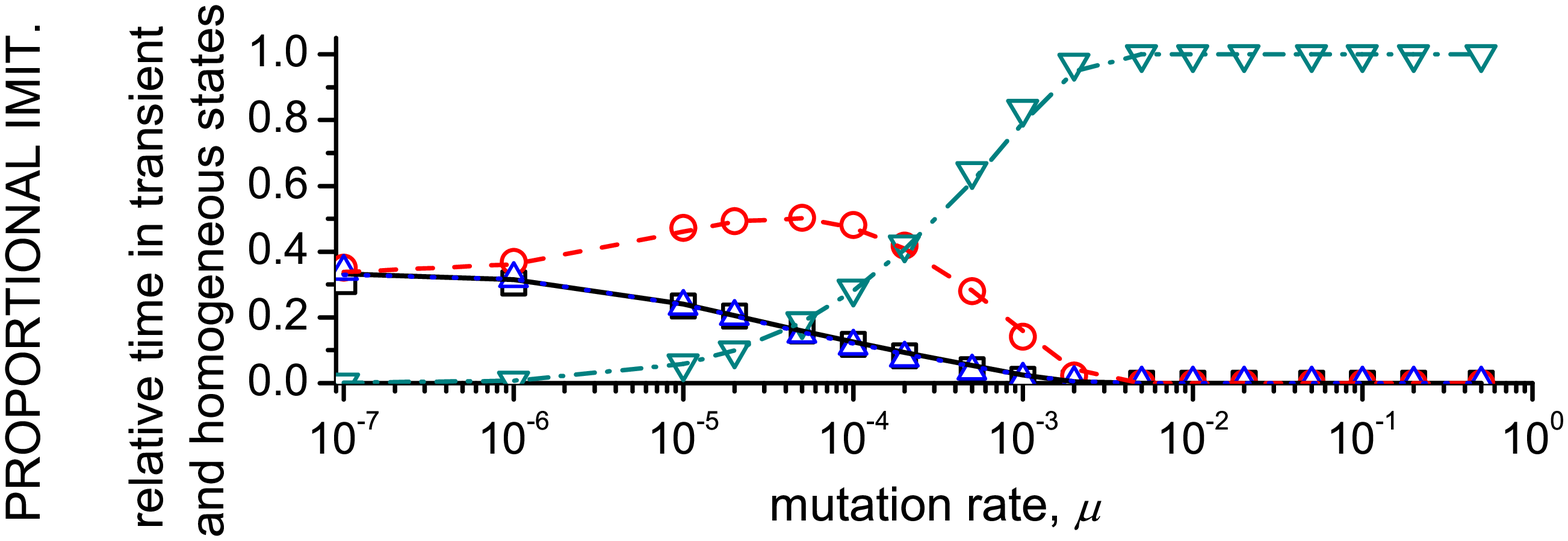} &
\includegraphics[width=88mm]{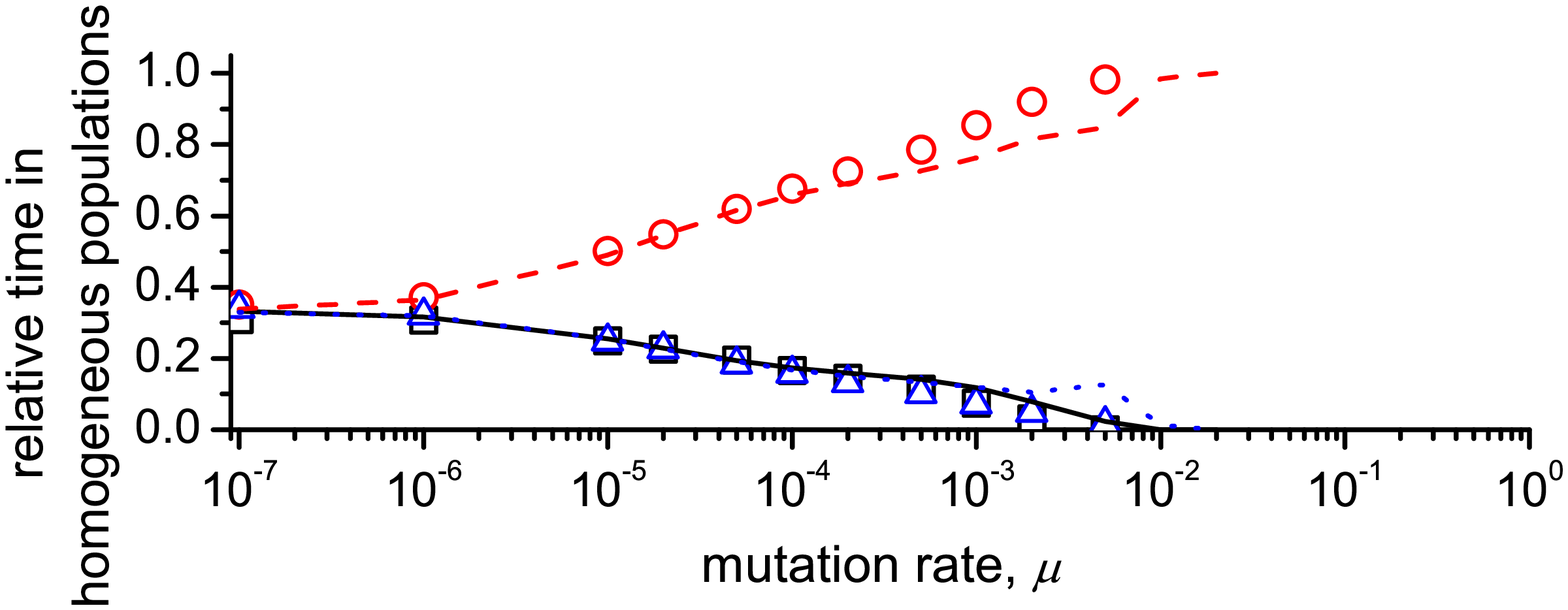} \\
(e) & (f) \\
\includegraphics[width=88mm]{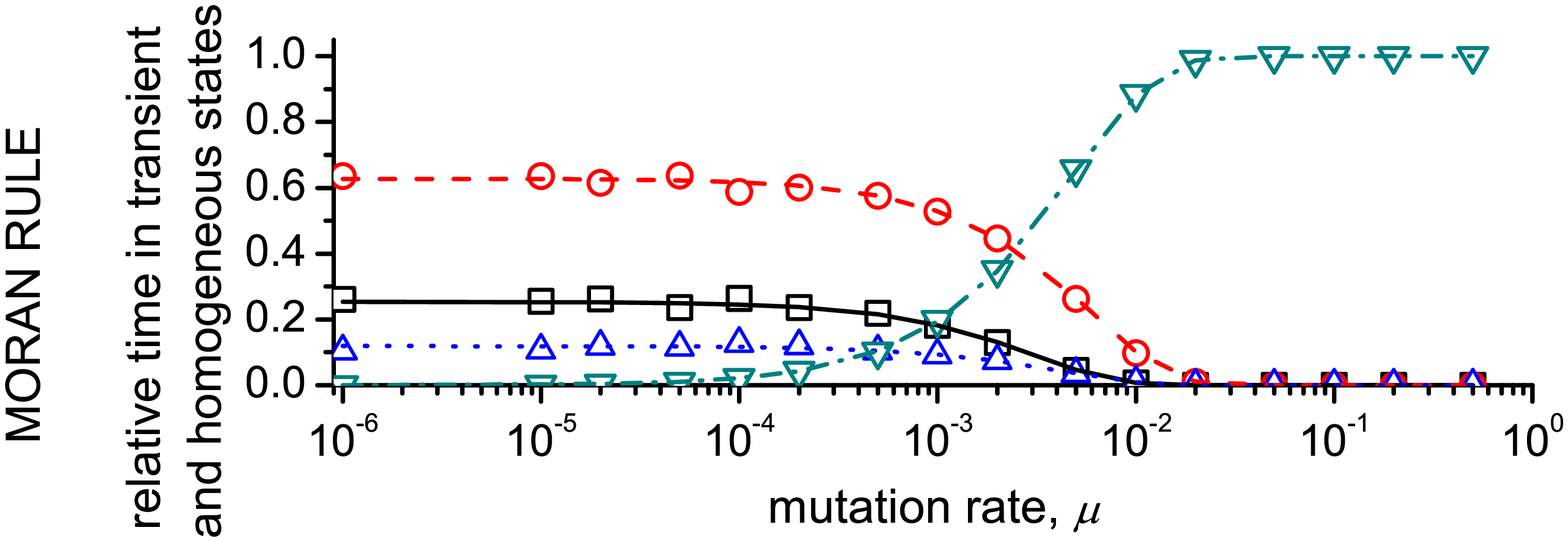} &
\includegraphics[width=88mm]{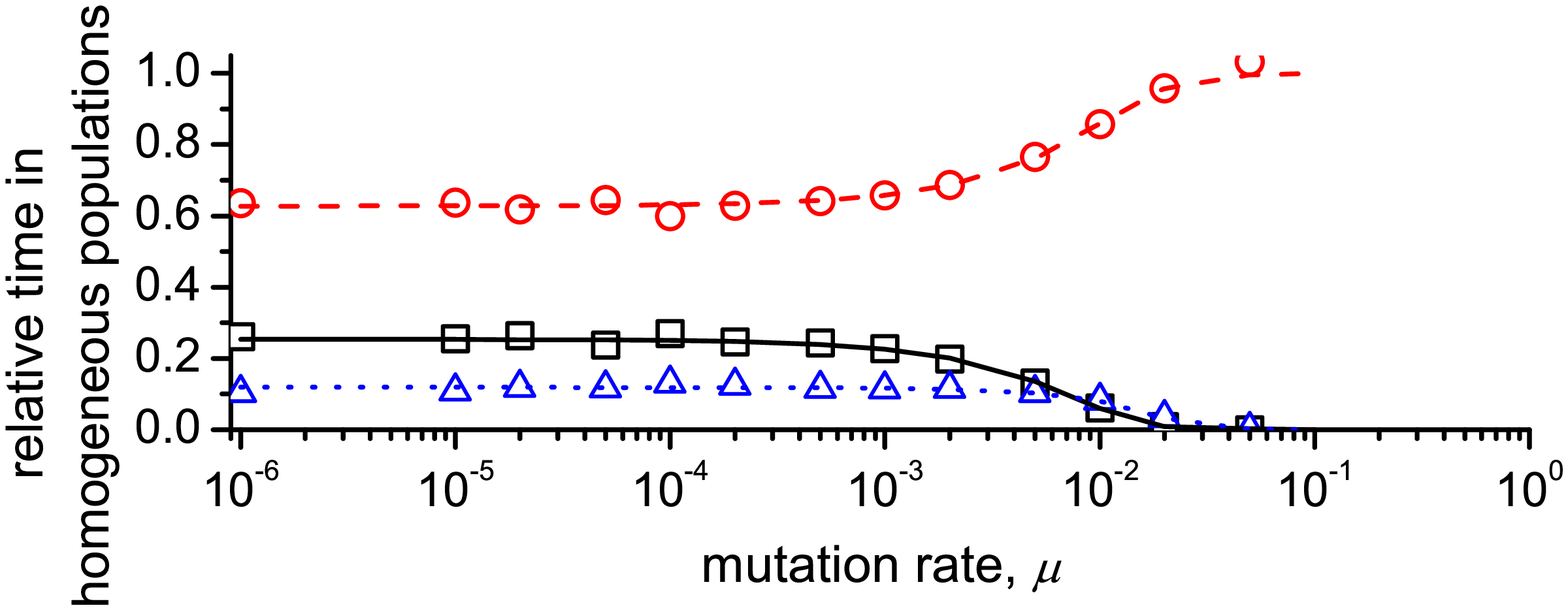}
\end{tabular}
\caption{(Color online.)
Relative times spent in homogeneous as well as in transient states
in a population of $M=100$ individuals. For practical purposes, a state is
considered homogeneous if more than 95\% of individuals belong to the same
strategy. Symbols are the result of agent-based simulations; lines are
obtained from the solution of Eqs.~\eqref{eq:T1}--\eqref{eq:pi1}. Results
for cooperators are represented with (black) squares and solid lines, those
for defectors with (red) circles and dashed lines, and those for jokers with
(blue) triangles and dotted lines. Panels (a), (c) and (e) also show (with
inverted triangles and dashed-dotted lines) the fraction of time spent in
transient states. Panels (b), (d) and (f) show the fractions of
the time spent in each of the three homogeneous states relative to the total
time spent in homogeneous states.
Panels (a) and (b) correspond to unconditional imitation, panels
(c) and (d) to proportional update, and panels (e) and (f) to a Moran rule.
We can see that high mutation rates promote defection over the other two
strategies. Parameters used are $n=5,r=3,d=0.4$; selection strength is
$s=1$ in (a)--(d) and $s=0.38$ in (e) and (f).}
\label{fig:freqvsmu}
\end{figure*}

Roughly speaking we can distinguish three regimes of mutations. In the low
mutation regime the system spends most of the time in homogeneous states,
and the dynamics of the system is well described by the $\mu\to 0$ limit
of the stationary probability distribution $\boldsymbol\pi$. This can be
clearly seen in Fig.~\ref{fig:freqvsmu}. The dashed-dotted curves in
Figs.~\ref{fig:freqvsmu}(a), (c) and (e) represent the fraction of time
spent in transients when a homogeneous population is replaced by another
one arisen as the result of mutations. This fraction is very small for
$\mu\lesssim 10^{-5}$--$10^{-4}$, depending on the imitation rule. For
larger mutation rates ($10^{-5}$--$10^{-4}\lesssim\mu\lesssim
10^{-3}$--$10^{-2}$) the system spends as much time in homogeneous populations
as in mixed transient states.
This is the regime displayed in Fig.~\ref{fig:updatecomp}, where cycles
are clearly defined even though for some imitation rules (particularly
so for proportional update) certain homogeneous populations that are hardly
ever reached [Fig.~\ref{fig:updatecomp}(b) shows burst of cooperators
which never reach a fraction higher than 80\% of the population]. For
even higher mutation rates homogeneous populations are very rare and
the behavior of the system is very different, typically dominated
by defectors [see Figs.\ref{fig:freqvsmu}(b), (d) and (f)].

\begin{figure}
\begin{center}
\includegraphics[width=88mm]{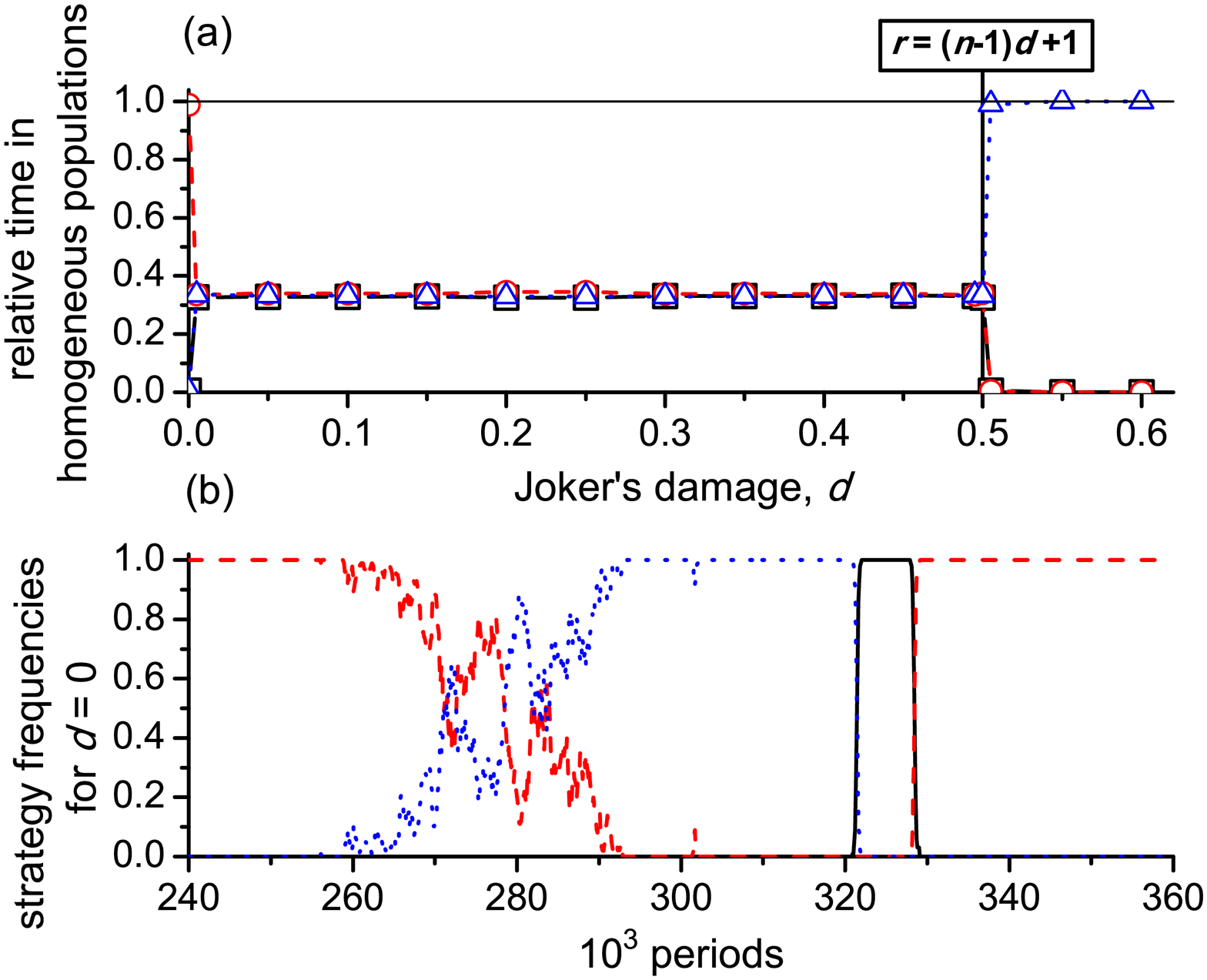}
\end{center}
\caption{(Color online.) Evolution of a population of $M=100$ individuals
by unconditional imitation. (a) Fraction of time spent in homogeneous
populations of cooperators (black squares and solid line), defectors (red
circles and dashed line) and jokers (blue triangles and dotted line),
as a function of joker's inflicted damage $d$. Symbols correspond
to an agent-based simulation; lines to the results obtained from
numerical computation of the stationary probability distribution.
(b) A realization made with $d=0$ showing an invasion of defectors
by jokers through pure drift, and the subsequent burst of cooperators
and turn-over by defectors. Parameters: $n=5,r=3,s=1$ and $\mu=5\times
10^{-5}$.}
\label{fig:uncimit}
\end{figure}

\begin{figure}
\begin{center}
\includegraphics[width=88mm]{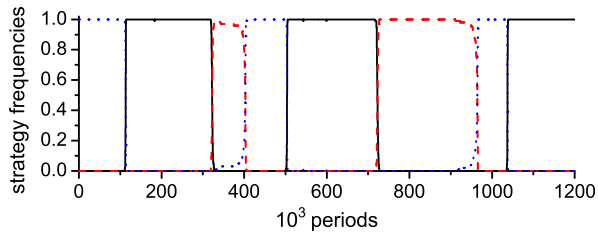}
\end{center}
\caption{(Color online.) Realization of an agent-based simulation of a
population with $M=100$ individuals evolving through proportional update.
Notation is as in Fig.~\ref{fig:uncimit}. 
Parameters: $n=5,r=3,d=0.4,\mu=5\times 10^{-6},s=1$.}
\label{fig:propimit}
\end{figure}

\begin{figure}
\begin{center}
\includegraphics[width=88mm]{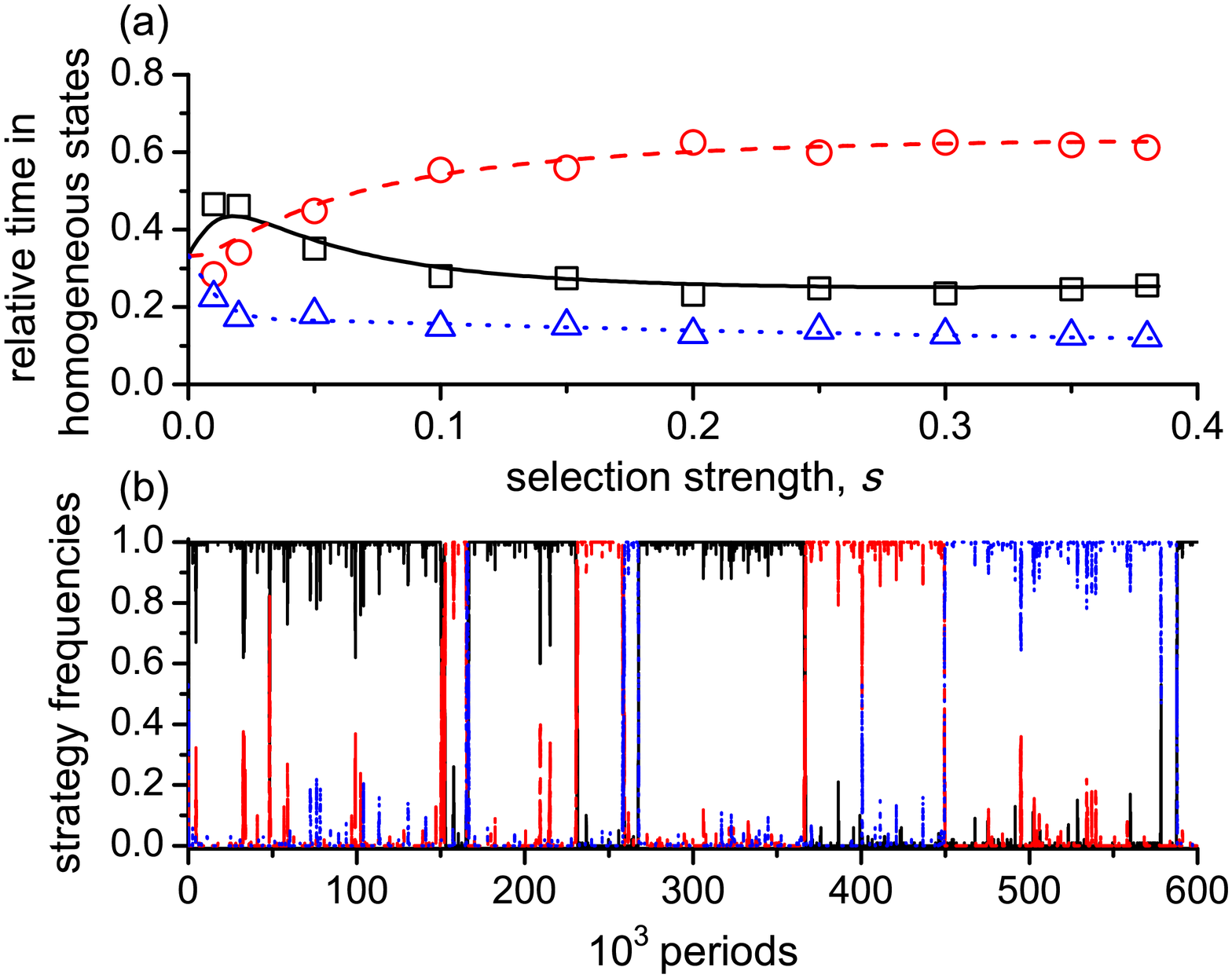}
\end{center}
\caption{(Color online.) A population of $M=100$ individuals evolving
through Moran update. (a) Comparison of the the relative times in which the
population is in a homogeneous state vs.\ the selection strength, $s$, for
low mutation rates. Lines represent the analytical estimates obtained in
Sec.~\ref{sec:appMoran}; symbols represent the results from agent-based
simulations. (b) Fractions of each strategy as a function of time as obtained
from a realization of an agent-based simulation. Cooperators are represented
with a black solid line and squares, defectors with a red dashed line and
circles, and jokers with a blue dotted line and triangles. Parameters are
$n=5,r=3,d=0.4$ and $\mu=5\times 10^{-5}$. In (b) the selection strength is
$s=0.05$.}
\label{fig:moran}
\end{figure}

Unconditional imitation is practically a deterministic rule in the low
mutations regime. For $\mu\lesssim 10^{-4}$ the population is almost
always homogeneous, and is made of each of the three strategies with
equal probability [see Figs.~\ref{fig:updatecomp}(a), (b) and
Figs.~\ref{fig:propimit}(a), (b)]. Figure~\ref{fig:freqvsmu}(a) shows
this probability as a function of the joker's inflicted damage $d$.
As long as $d>0$ and $r>1+(n-1)d$ we find each strategy equally likely.
For $r<1+(n-1)d$ a homogeneous population of jokers cannot be invaded
because this is the only absorbing state of the Markov chain for $\mu=0$.
For $d=0$ jokers do not inflict damage. Then the system spends most
of the time in a homogeneous population of defectors. However, random drift
allows for occasional invasions by jokers, who are subsequently wiped
out by cooperators, who in its turn get replaced again by defectors.
Figure~\ref{fig:uncimit}(b) illustrates a typical realization exhibiting
one of these turn-overs.

As of proportional update, its main difference with unconditional imitation is
its being a truly probabilistic rule, in which individuals only imitate higher
payoffs with a certain probability. Although in the small mutations regime
this leads to the same probability of mutual invasion of strategies as for
unconditional imitation, the stochastic nature of this rule renders much
longer invasion times. This can be clearly appreciated in
Fig.~\ref{fig:updatecomp}.

Another effect of stochasticity is that the time spent in transient states
is also longer, thus shrinking the low mutations regime by more than one
order of magnitude [compare Figs.~\ref{fig:freqvsmu}(a) and (c)]. The
effect is particularly notorious for jokers, who take a long time to
invade defectors, thus extending the life time of defective populations.
This effect is illustrated in Fig.~\ref{fig:propimit}, which represents a
typical realization of an agent-based simulation.

The Moran process is the randomest of the three evolutionary dynamics
because even strategies not performing very well have a chance to get imitated.
The effect is more noticeable the smaller the population. This dynamics
imposes an upper limit to the selection strength $s$ (see
Sec.~\ref{sec:imitation}) and the probabilities to find the population in each
of the three homogeneous states depend on the parameters of the game and on
$s$ in a nontrivial way (see Sec.~\ref{sec:appMoran}). These probabilities are
represented in Fig.~\ref{fig:moran}(a) as a function of $s$. The theoretical
predictions of Sec.~\ref{sec:appMoran} agree with the simulations. This figure
shows that cooperation is highly promoted for small $s$($0.005<s<0.05$). In
this limit cooperative populations are found with almost 50\% probability. This
probability decreases down to around 25\% for larger $s$.
Figure~\ref{fig:moran}(b) shows a typical realization of this process,
exhibiting a defining feature of this process, namely the frequent failures of
attempted invasions.

\begin{figure*}
\begin{tabular}[t]{ccc}
(a) & (b) & (c)\\
\includegraphics[width=55mm]{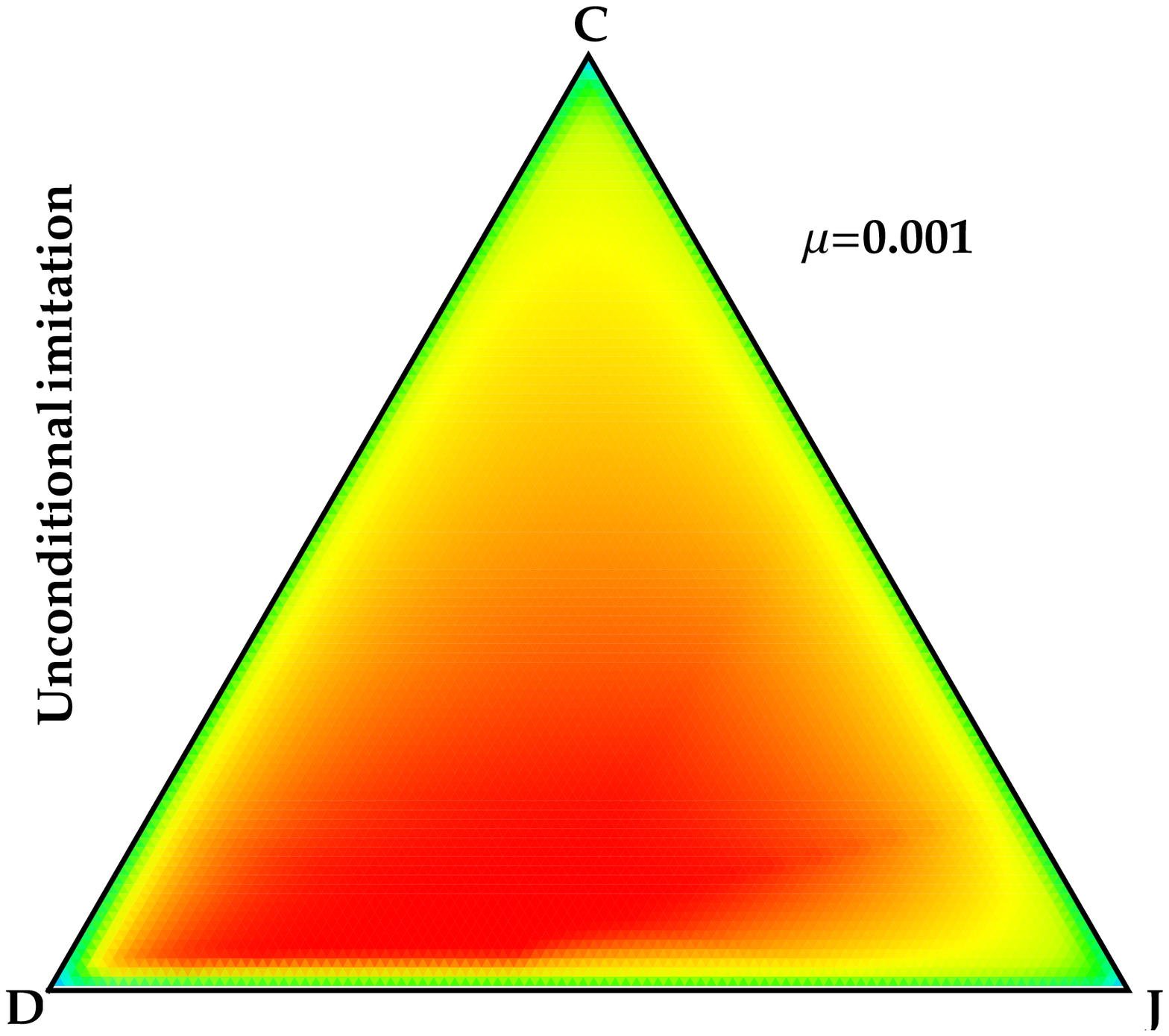} &
\includegraphics[width=55mm]{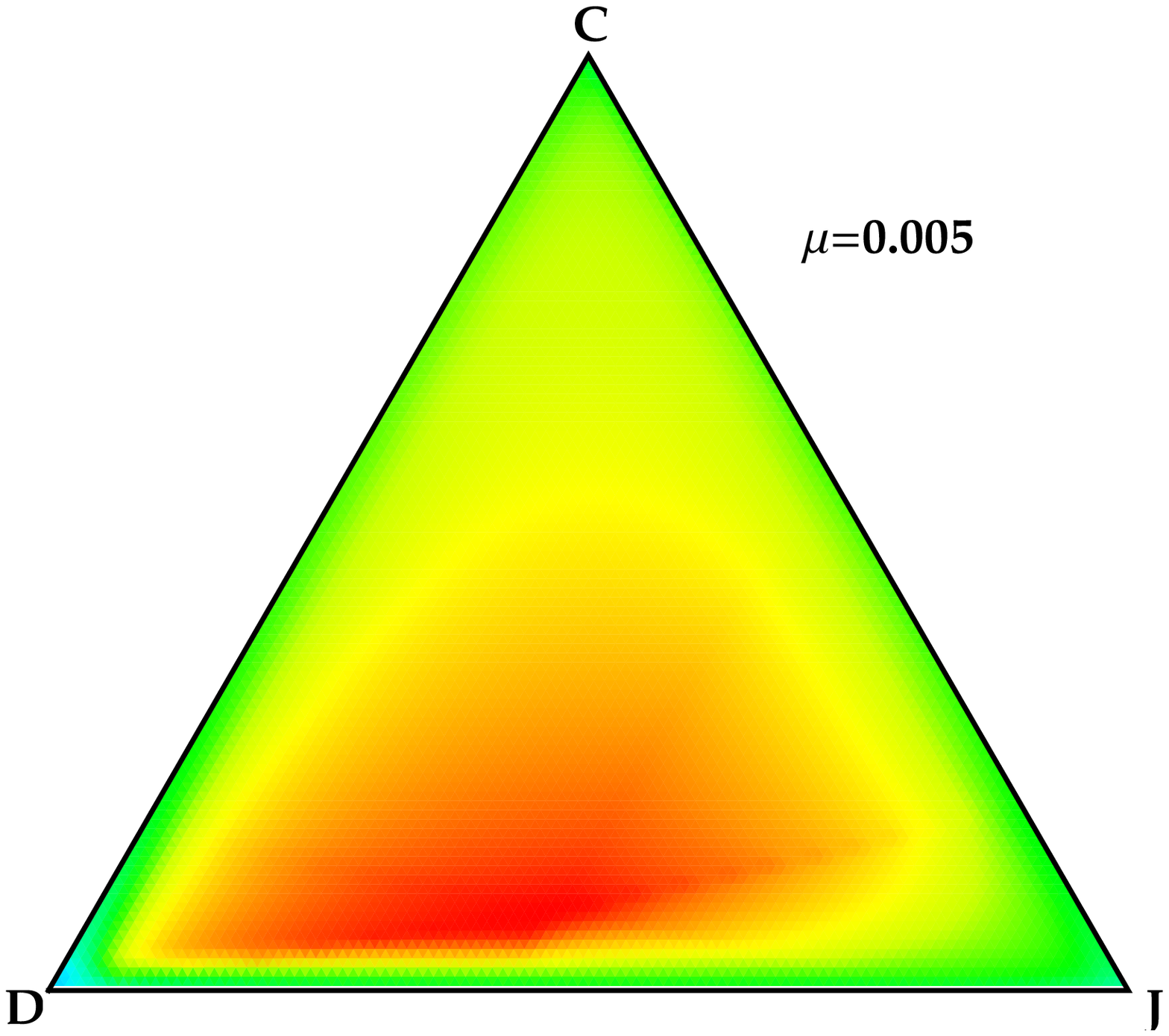} &
\includegraphics[width=55mm]{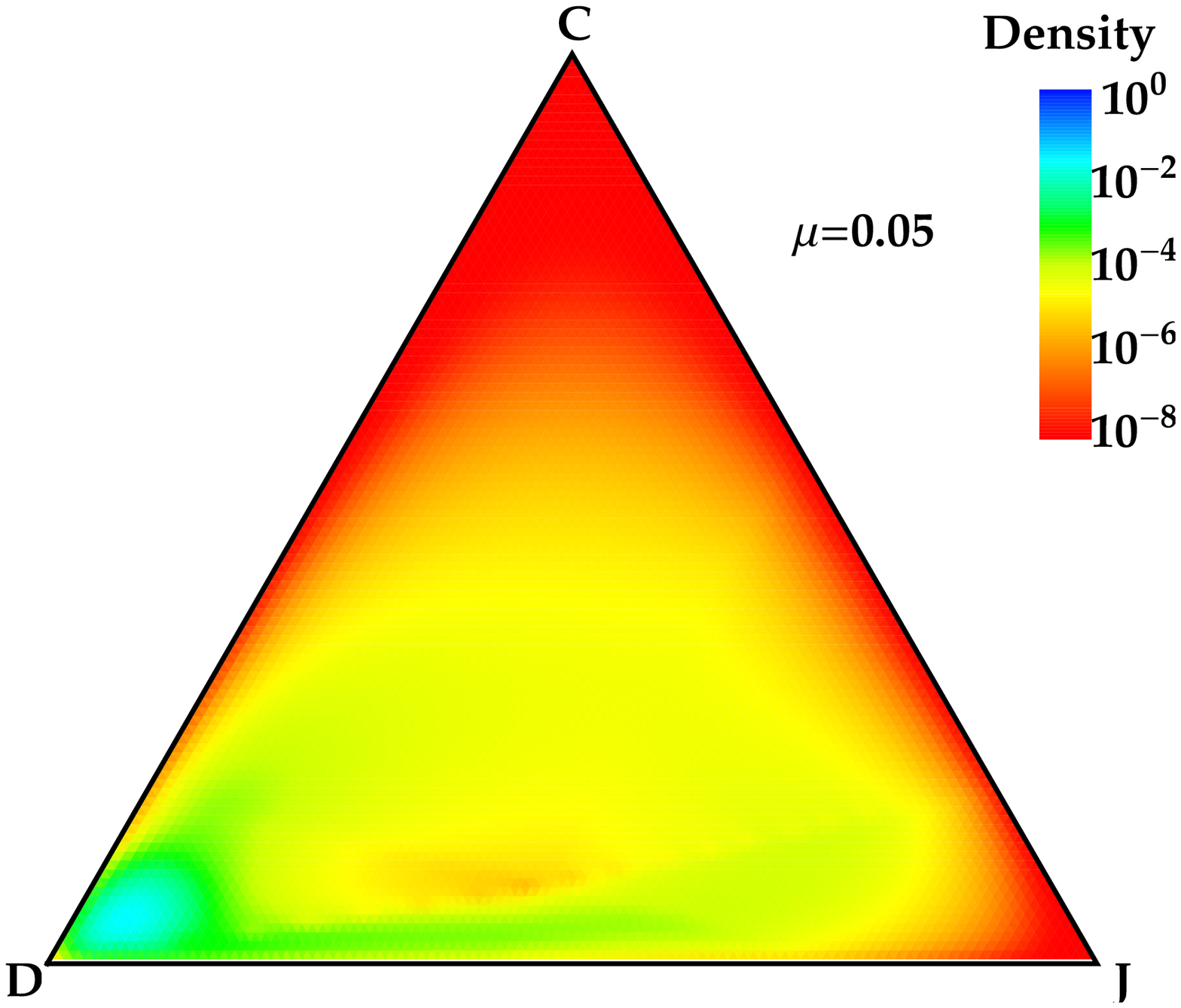} \\
(d) & (e) & (f) \\
\includegraphics[width=55mm]{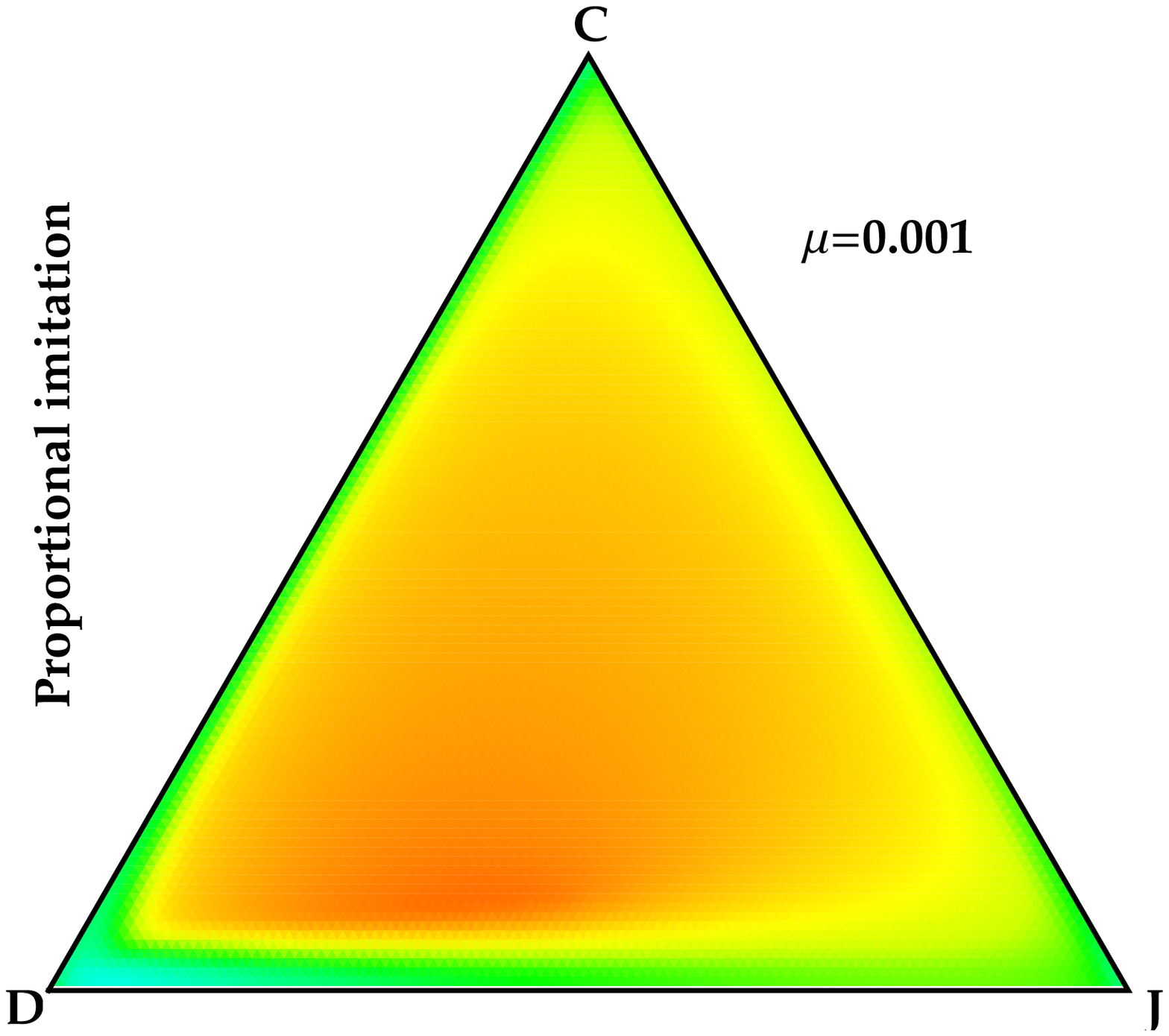} &
\includegraphics[width=55mm]{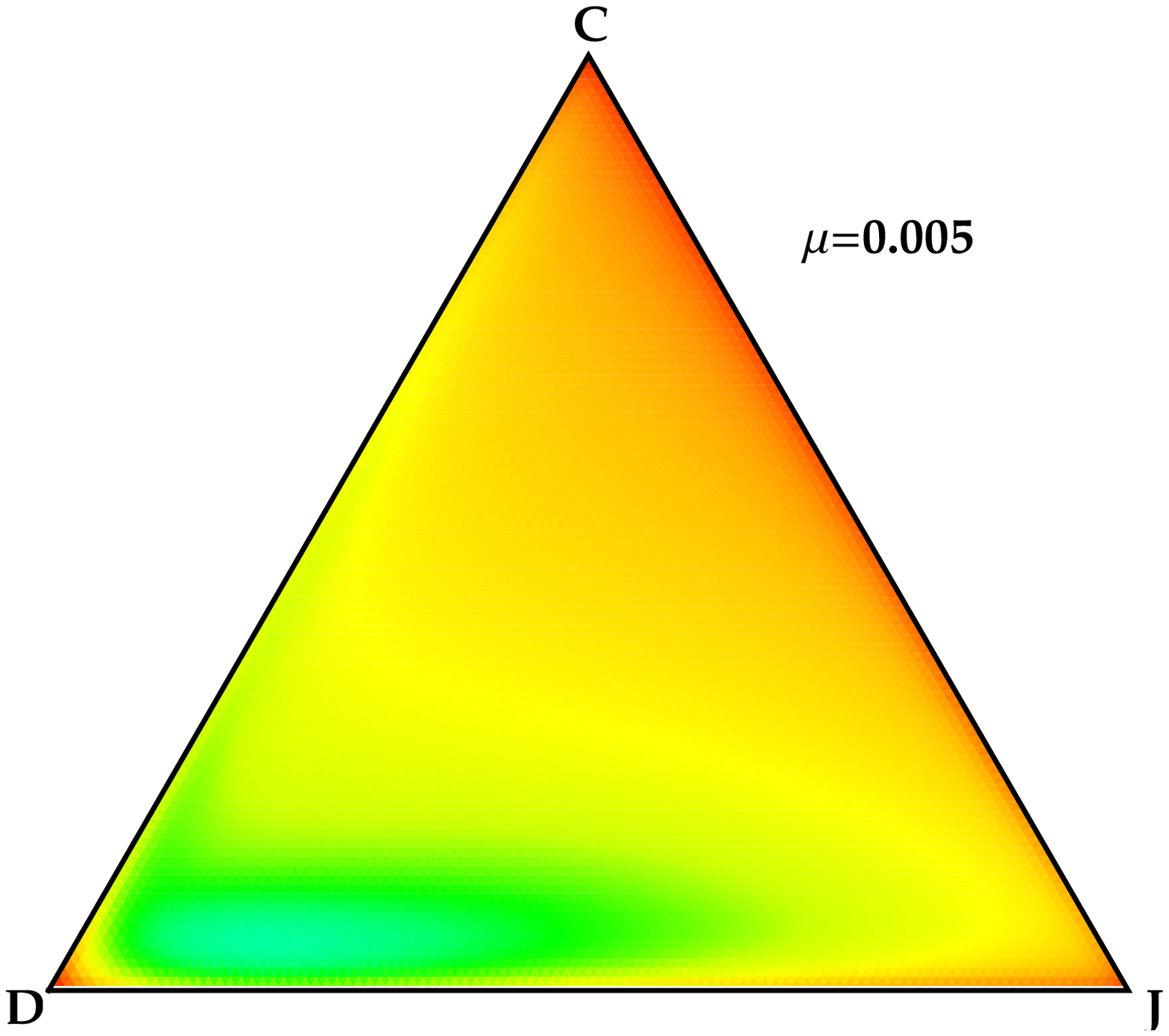} &
\includegraphics[width=55mm]{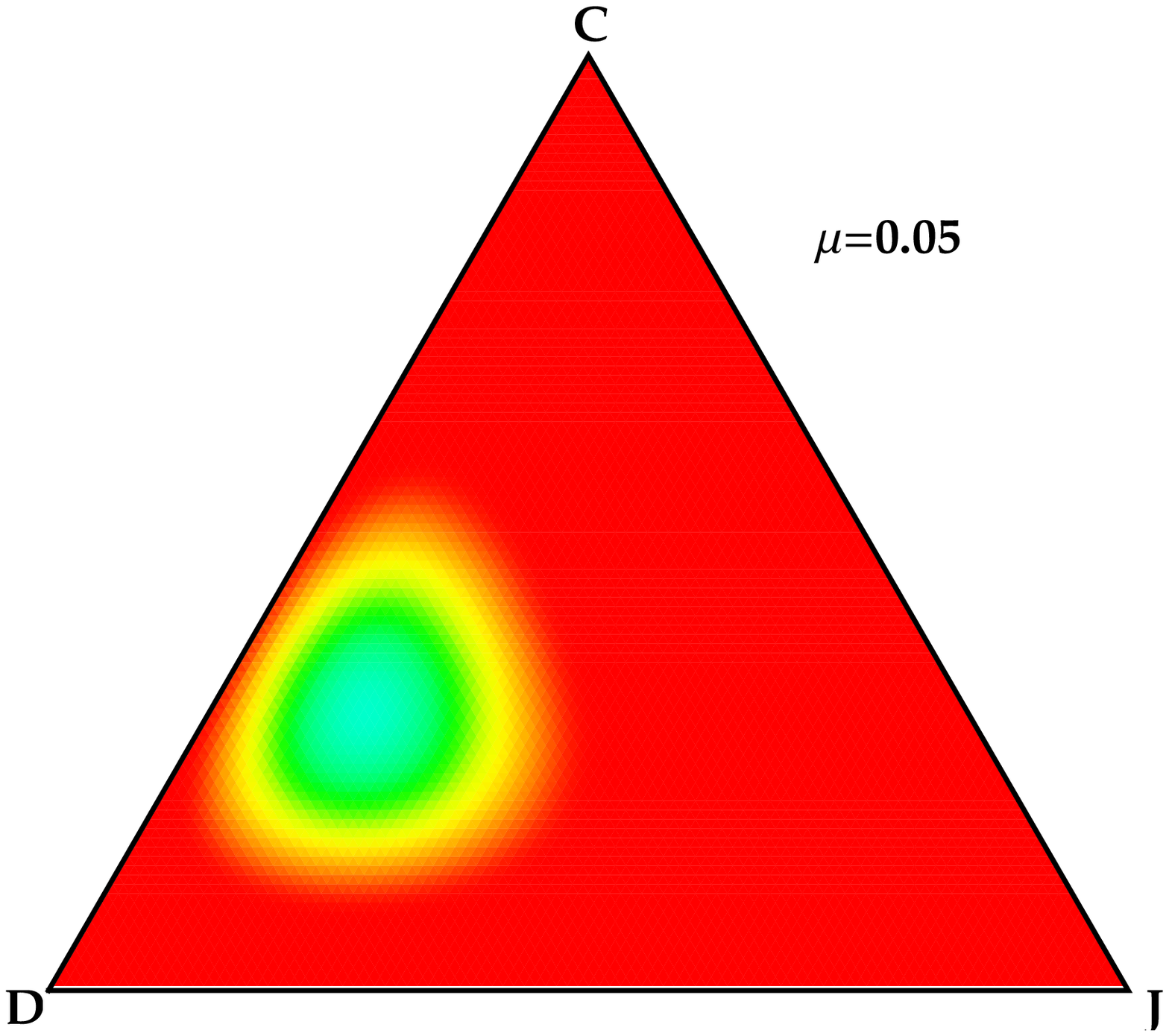} \\
(g) & (h) & (i) \\
\includegraphics[width=55mm]{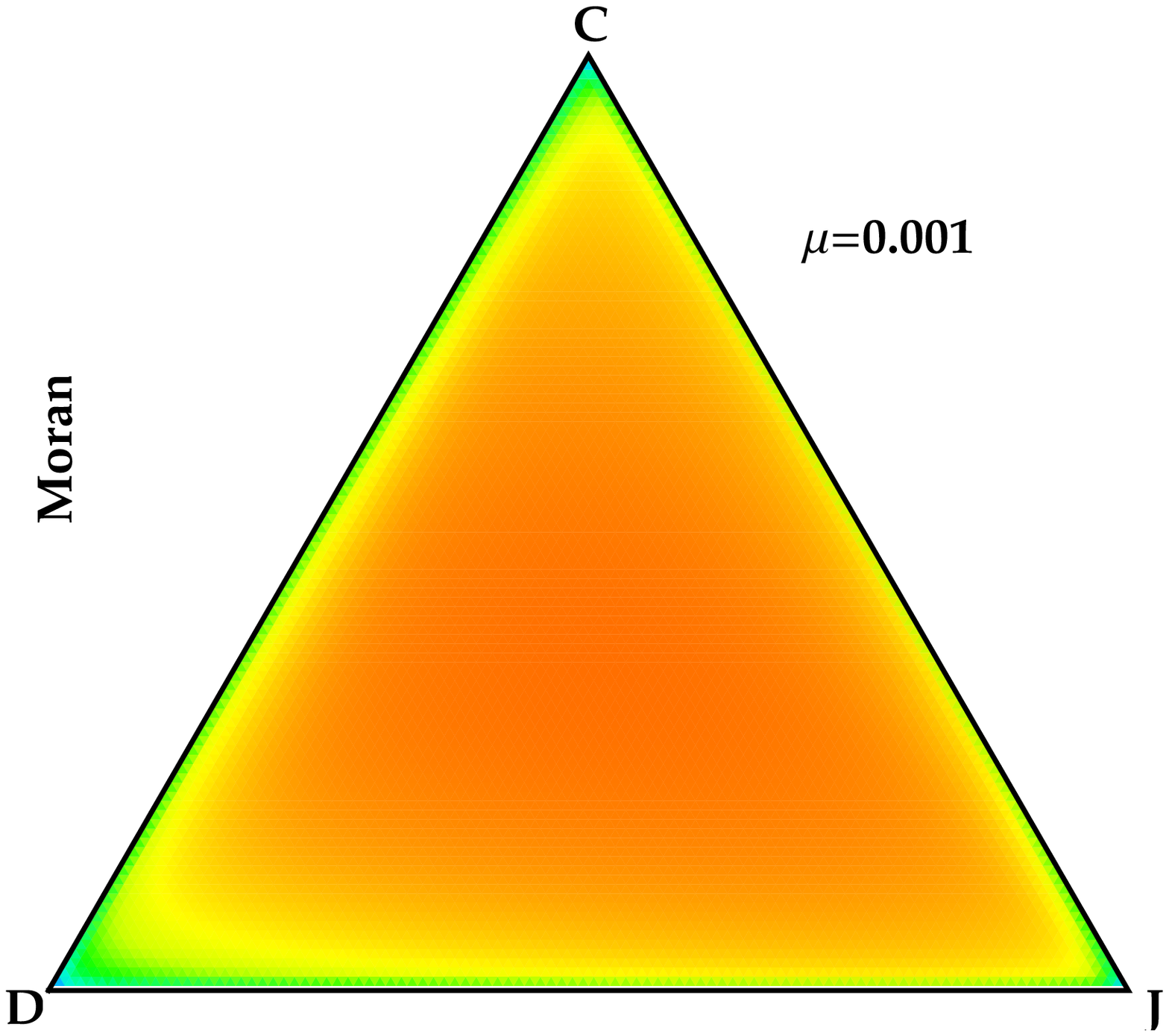} &
\includegraphics[width=55mm]{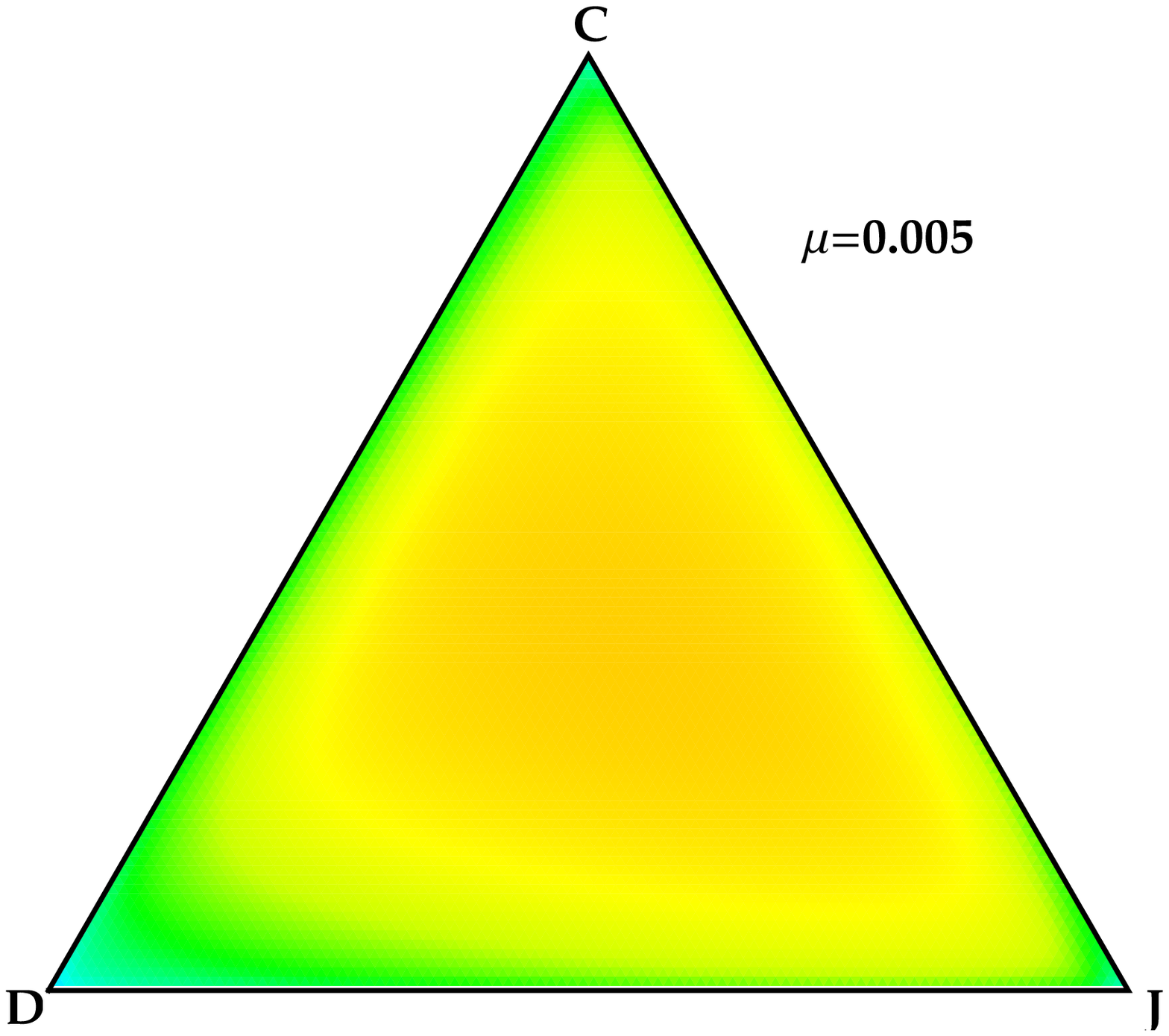} &
\includegraphics[width=55mm]{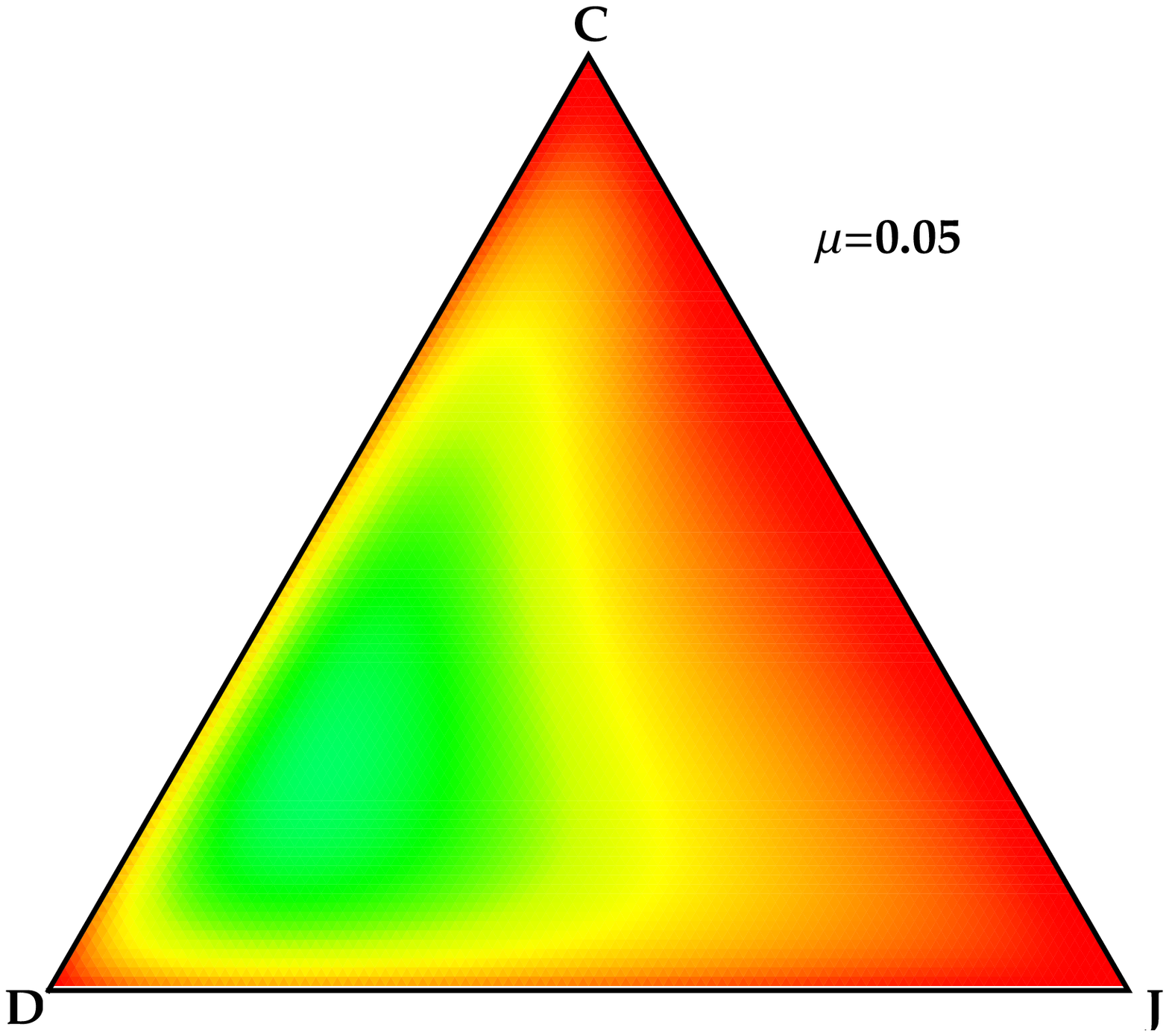} \\
\end{tabular}
\caption{(Color online.) Density plots for the probability of finding the system in each
population state as obtained by solving numerically Eq.~\eqref{eq:pistat}.
First row corresponds to unconditional imitation, second row to proportional
imitation, third row to a Moran process. In each case mutations increase left
to right.  In all three cases low mutation rates ($\mu$) yield high
probabilities near the boundaries of the simplexes, specially near the corners,
corresponding to cyclic transitions between homogeneous states. Increasing
$\mu$ increases the probability to find the system near homogeneous defective
populations. For high $\mu$ an attractive point appears close to the D corner
which goes away from it upon increasing $\mu$. Parameters are $n=5,r=3,d=0.4$;
the selection strength is $s=1$ in the first and second rows, $s=0.38$ in the
third. Mutation rates have been chosen as in Fig.~\ref{fig:simplex} and appear
near each simplex. Densities are plotted using a logarithmic scale.}
\label{fig:estoc}
\end{figure*}

Whichever the update rule, when mutation rates are not small the system is
better characterized by providing the stationary probability distribution
$\pi$, as obtained from Eq.~\eqref{eq:pi1}). The results are plotted in
Fig.~\ref{fig:estoc} for all three imitation rules and different mutation
rates $\mu$. For low and intermediate values of $\mu$ the higher probabilities
are found near the border of the simplexes, consistent with the cyclic behavior
of the system. However, for high $\mu$ the probability peaks around a point.
This point is interior for the most stochastic rules, but corresponds to
a defective population for unconditional imitation. The simplexes are obtained
for the same parameter values as used in Fig.~\ref{fig:simplex}, so a direct
comparison with the results of the replicator dynamics can be made.

\section{Discussion and conclusions}

In this paper we have proven that the existence of jokers, i.e., individuals
whose purely destructive behavior is directed against the common enterprises
represented by PG games, allows for the emergence of robust evolutionary cycles
in finite populations regardless of the updating method chosen. Together with a
previous report \cite{arenas:2011} on the existence of limit cycles for
infinite populations evolving via a replicator-mutator dynamics, our present
results show that limit cycles are a generic feature of the dynamics generated
by destructive agents, not restricted to a particular selection dynamics. In
fact, this is a dynamical feature that makes this model different from other
three-player games like that of loners \citep{hauert:2002b}, for which cycles
are structurally unstable and their existence strongly depends on the absence
of mutations and other kinds of perturbations.

In a recent paper \citep{mobilia:2010} Mobilia has shown that, of the three
possible outcomes of the replicator equation for rock-paper-scissors games
\cite{hofbauer:1998}, namely (a) orbits are attracted towards an asymptotically
stable mixed equilibrium, (b) orbits cycle around a neutrally stable mixed
equilibrium, and (c) orbits go away from an unstable mixed equilibrium and
approach the heteroclinic orbit defined by the border of the simplex (the
case of the Joker game), adding mutations between the three strategies
merges cases (a) and (b), both of which yielding a mixed equilibrium.
Oscillations disappear in these two cases. The Loners game belongs to 
class (b). In contrast, the Joker case analysed in this paper belongs to class (c), 
with mutations generating an attractive, stable limit
cycle. In this case the dynamics oscillates between the three strategies with
well-defined and robust oscillations. We
are not aware of any other game for which the inclusion of a simple behavioral
type (jokers do not need memory, have no especial recognition abilities, and
do not rely on any reputation generated along the game) leads to cycles which
are robust to perturbations and have a well defined period and amplitude
irrespective of the initial fractions of players. 

We have expanded here these results and have proven that the oscillatory
dynamics does not occur only for infinite (or very large) populations evolving
under a replicator dynamics, but also in the case of finite populations and for
different update rules. We have analyzed unconditional imitation, proportional
update, and a Moran process. In all cases the system exhibits finite time
lapses in which most of the population is composed by cooperative individuals,
finding that the Moran process for low (but not extremely low) selection
pressures is the most favorable to cooperation---as the system spends 50\% of
the time in cooperative states. Under unconditional imitation the system spends one third of the time in cooperative states, whereas the more stochastic nature of
proportional update favors defection due to the slower invasion of jokers, and
thus the system stays longer in defective states---especially so for high
mutation rates.

Let us note that, if the damage $d$ inflicted by jokers is zero, jokers are not able to
overcome defectors and oscillations are supressed. The system ends up in a steady population where cooperation becomes extinguished, both with and without mutations \citep{arenas:2011}. Indeed, this case is identical to the loner model when the benefit obtained by loners is also zero, situation in which both jokers and loners become simply non-participants in the game with the only effect of reducing the effective number of players \citep{hauert:2002b,arenas:2011}. We have shown here that for finite populations and $d=0$ random drift allows for bursts in which the system spends some time in fully cooperative states, but that the happening probability of such events is very low.

The existence of damaging agents, which are able to destroy the defective
populations and lead to a state without cooperators and defectors,
gives cooperators the chance to re-build cooperative enterprises, and
thus promotes cooperation. This result, as well as modifications of the
model presented here, might be interesting in the study of human
evolution, where examples of destructive periods can be found along history as
a result of revolutions or wars. It has been suggested that these destructive
periods take place whenever a society reaches a point where the public goods fall
below a certain threshold \citep{turchin:2006}. The Joker game shares this
feature. Modifications of the model presented here may thus help explain not
only how cooperation in animals arises whenever there is a risk or they face a
predator, but also provide insights into the evolutionary cycles observed
in human society.

\section*{Acknowledgments} Financial support from Ministerio de Ciencia y
Tecnolog\'{\i}a (Spain) under projects FIS2009-13730-C02-02 (A.A.),
FIS2009-13370-C02-01 (J.C. and R.J.R.), MOSAICO, PRODIEVO and Complexity-NET
RESINEE (J.A.C.); from the Barcelona Graduate School of Economics and
of the Government of Catalonia (A.A.); from the Generalitat de Catalunya under
project 2009SGR0838 (A.A.) 2009SGR0164 (J.C. and R.J.R.) and from Comunidad de
Madrid under project MODELICO-CM (J.A.C.). R.J.R. acknowledges the financial
support of the Universitat Aut\`onoma de Barcelona (PIF grant) and the Spanish
government (FPU grant).

%% The Appendices part is started with the command \appendix;
%% appendix sections are then done as normal sections
\appendix

\section{Average payoffs in a finite population}
\label{app:B}

Let us denote $P_{\rm X}(m,l)$ the
average payoff that a player of type X receives when the population is
made of $m$ cooperators, $j$ jokers, and $M-m-j$ defectors. This average
payoff is calculated by averaging the corresponding payoff \eqref{eq:payoffs}
with the probability distribution \eqref{eq:general-hyper}. For defectors
this implies
\begin{equation}
P_{\rm D}(m,j) = \sum_{\substack{k,l\ge 0 \\ k+l<n}}\frac{rk-dl}{n-l}
p(k,l|n-1,m,j,M-1).
\label{eq:averageD}
\end{equation}
To perform this average it will prove convenient to factorize the
probability distribution as the product of two standard hypergeometric
distributions, i.e.,
\begin{equation}
p(k,l|n,m,j,M)=p(l|n,j,M)p(k|n-l,m,M-j).
\label{eq:factorize}
\end{equation}
where
\begin{equation}
p(l|n,j,M)=\frac{\displaystyle \binom{j}{l}\binom{M-j}{n-l}}{\displaystyle
\binom{M}{n}}.
\label{eq:simple-hyper}
\end{equation}
The first term in \eqref{eq:factorize} is the probability of selecting $l$
jokers out of the population, and the second term is the conditional
probability of subsequently selecting $k$ cooperators, given that we have
already selected the $l$ jokers.

A useful identity of the hypergeometric distribution ---consequence of
the properties of the binomial coefficients--- is
\begin{equation}
k\,p(k|n,m,M)=\frac{nm}{M}p(k-1|n-1,m-1,M-1).
\end{equation}
Substituting factorization \eqref{eq:factorize} into \eqref{eq:averageD}
and making use of this identity we readily obtain
\begin{equation}
\begin{split}
P_{\rm D}(m,j)
=&\, \frac{rm}{M-j-1}\sum_{l=0}^{n-1}\frac{n-l-1}{n-l}\,p(l|n-1,j,M-1) \\
&-d\sum_{l=0}^{n-1}\frac{l}{n-l}\,p(l|n-1,j,M-1).
\end{split}
\label{eq:PDint}
\end{equation}

A new identity, namely
\begin{equation}
\frac{p(l|n-1,j,M-1)}{n-l}=\frac{M}{n(M-j)}p(l|n,j,M),
\end{equation}
allows us to do the sum
\begin{equation}
\sum_{l=0}^{n-1}\frac{p(l|n-1,j,M-1)}{n-l}=\frac{M[1-p(n|n,j,M)]}{n(M-j)}.
\end{equation}
It will prove convenient to introduce the function
\begin{equation}
\Xi(n,j,M)\equiv \frac{j}{M-j}\left[1-
\frac{(j-1)\cdots(j-n+1)}{(M-1)\cdots(M-n+1)}\right],
\end{equation}
in terms of which
\begin{equation}
\begin{split}
1-p(n|n,j,M) &=1-\frac{j(j-1)\cdots(j-n+1)}{M(M-1)\cdots(M-n+1)} \\
&=\frac{M-j}{M}\left[1+\Xi(n,j,M)\right].
\end{split}
\end{equation}
This allows us to write
\begin{equation}
\sum_{l=0}^{n-1}\frac{p(l|n-1,j,M-1)}{n-l}=\frac{1+\Xi(n,j,M)}{n},
\end{equation}
and using this in~\eqref{eq:PDint} obtain
\begin{equation}
P_{\rm D}(m,j) =\frac{rm[n-1-\Xi(n,j,M)]}{n(M-j-1)}-d\,\Xi(n,j,M).
\end{equation}

As for the average payoff of a cooperator,
\begin{equation}
\begin{split}
P_{\rm C}(m,j) &= -1+\sum_{\substack{ k,l\ge 0 \\ k+l<n}}\frac{r(k+1)-dl}{n-l} \\
&\phantom{=}\times
p(k,l|n-1,m-1,j,M-1) \\
&= r\sum_{l=0}^{n-1}\frac{p(l|n-1,j,M-1)}{n-l}-1
+P_{\rm D}(m-1,j) \\
&= \frac{r}{n}[1+\Xi(n,j,M)]-1+P_{\rm D}(m-1,j).
\end{split}
\label{eq:averageC}
\end{equation}
Therefore
\begin{equation}
\begin{split}
P_{\rm C}(m,j) =&\,\frac{r}{n}\left(1+\frac{(n-1)(m-1)}{M-j-1}\right)-1 \\
&+\left[\frac{r}{n}\left(1-\frac{m-1}{M-j-1}\right)-d\right]\Xi(n,j,M).
\end{split}
\end{equation}

Finally, $P_{\rm J}(n,j)=0$ because jokers get zero regardless of
the composition of the population.

\section{Calculation of the transition matrices}
\label{app:C}

Transition probabilities $T(m,j|m',j')$ are obtained according to
the specified update rule. We will calculate those corresponding
to the rules used in this work. But before we proceed let us
introduce some shorthands. We will write $T_{\epsilon_1,\epsilon_2}
\equiv T(m,j|m+\epsilon_1,j+\epsilon_2)$, where $\epsilon_1,\epsilon_2
\in\{-1,0,1\}$. Also by $\omega^{\rm XY}_{\epsilon_1,\epsilon_2}$
we will denote the probability that a player of type Y is chosen
to be replaced by a player of type
X when the population is made of $m+\epsilon_1$ cooperators,
$j+\epsilon_2$ jokers, and $M-m-j-\epsilon_1-\epsilon_2$ defectors.
Whether the Y player is finally replaced by an X one depends on
mutations, thus
\begin{equation}
\begin{split}
T_{1,0}=&\, \omega^{\rm DC}_{1,0}(1-2\mu)
+\left(\omega^{\rm JC}_{1,0}+\omega^{\rm CC}_{1,0}\right)\mu, \\
T_{1,-1}=&\, \omega^{\rm JC}_{1,-1}(1-2\mu)
+\left(\omega^{\rm DC}_{1,-1}+\omega^{\rm CC}_{1,-1}\right)\mu, \\
T_{-1,0}=&\, \omega^{\rm CD}_{-1,0}(1-2\mu)
+\left(\omega^{\rm JD}_{-1,0}+\omega^{\rm DD}_{-1,0}\right)\mu, \\
T_{0,-1}=&\, \omega^{\rm JD}_{0,-1}(1-2\mu)
+\left(\omega^{\rm CD}_{0,-1}+\omega^{\rm DD}_{0,-1}\right)\mu, \\
T_{-1,1}=&\, \omega^{\rm CJ}_{-1,1}(1-2\mu)
+\left(\omega^{\rm DJ}_{-1,1}+\omega^{\rm JJ}_{-1,1}\right)\mu, \\
T_{0,1}=&\, \omega^{\rm DJ}_{0,1}(1-2\mu)
+\left(\omega^{\rm CJ}_{0,1}+\omega^{\rm JJ}_{0,1}\right)\mu,
\end{split}
\label{eq:Ts}
\end{equation}
In all cases there are two possibilities for a Y individual to become
an X one, either a pair XY is selected, the update takes place and no
mutation occurs, or another pair ZY is selected (with ${\rm Z}\ne{\rm X}$)
but a mutation changes Z into X.

Finally, the probability that no change of strategy occurs $T_{0,0}=
T(m,j|m,j)$ is obtained as
\begin{equation}
T_{0,0}=1-(1-\mu)
\sum_{{\rm X}\ne{\rm Y}}\omega^{\rm XY}_{0,0}
-2\mu\sum_{{\rm X}}\omega^{\rm XX}_{0,0},
\label{eq:T00}
\end{equation}
where the subscript $0,0$ refers to a population made of $m$ cooperators,
$j$ jokers, and $M-m-j$ defectors.

Notice that the expansion \eqref{eq:T1} readily follows from expressions
\eqref{eq:Ts} and \eqref{eq:T00}.

\subsection{Unconditional imitation}

This rule prescribes that two players are selected at random from the
population and the strategy of the model player (X) replaces that of
the focal player (Y) if the latter has a lower payoff. Accordingly, if ${\rm
X}\ne{\rm Y}$,
\begin{equation}
\omega^{\rm XY}_{\epsilon_1,\epsilon_2}=
\frac{n^{\rm X}_{\epsilon_1,\epsilon_2}
n^{\rm Y}_{\epsilon_1,\epsilon_2}}{M(M-1)}\Theta\left(
P^{\rm X}_{\epsilon_1,\epsilon_2}-P^{\rm Y}_{\epsilon_1,\epsilon_2}\right),
\end{equation}
where $\Theta(x)=1$ if $x>0$ and $0$ otherwise, and
$n^{\rm X}_{\epsilon_1,\epsilon_2}$ denotes the number of individuals of
type X in the population (e.g., $n^{\rm C}_{1,0}=m+1$, $n^{\rm D}_{1,0}=
M-j-m-1$, $n^{\rm J}_{1,-1}=j-1$, $n^{\rm D}_{0,0}=M-j-m$, etc.). On the
other hand, in order to account for mutations we must define
\begin{equation}
\omega^{\rm XX}_{\epsilon_1,\epsilon_2}=
\frac{n^{\rm X}_{\epsilon_1,\epsilon_2}
(n^{\rm X}_{\epsilon_1,\epsilon_2}-1)}{M(M-1)}.
\label{eq:wXX}
\end{equation}

\subsection{Proportional update}

Similarly to the previous rule,
\begin{equation}
\omega^{\rm XY}_{\epsilon_1,\epsilon_2}=
\frac{n^{\rm X}_{\epsilon_1,\epsilon_2}
n^{\rm Y}_{\epsilon_1,\epsilon_2}}{M(M-1)}\Psi\left(
P^{\rm X}_{\epsilon_1,\epsilon_2}-P^{\rm Y}_{\epsilon_1,\epsilon_2}\right),
\end{equation}
where $\Psi(x)=x/\Omega$ if $x>0$ and $0$ otherwise, $\Omega$ being
a constant ensuring that $\Psi\left(P^{\rm X}_{\epsilon_1,\epsilon_2}
-P^{\rm Y}_{\epsilon_1,\epsilon_2}\right)\le 1$ (typically $\Omega$
is chosen as the largest possible payoff difference).
As in the previous rule $\omega^{\rm XX}_{\epsilon_1,\epsilon_2}$ is given by
\eqref{eq:wXX}.

\subsection{Moran process}

In this case payoffs are replaced by fitnesses
$F^{\rm X}_{\epsilon_1,\epsilon_2}=1-s+sP^{\rm X}_{\epsilon_1,\epsilon_2}$
(see Sec.~\ref{sec:imitation}). Let us introduce the total fitness of the
population
\begin{equation}
\Phi_{\epsilon_1,\epsilon_2}\equiv\sum_{\rm X}
n^{\rm X}_{\epsilon_1,\epsilon_2}F^{\rm X}_{\epsilon_1,\epsilon_2}.
\end{equation}
The Moran rule specifies that a player is chosen for reproduction
proportional to its fitness and the offspring replaces another
 randomly chosen individual from 
the rest of the population. So if ${\rm X}\ne{\rm Y}$,
\begin{equation}
\omega^{\rm XY}_{\epsilon_1,\epsilon_2}=
\frac{n^{\rm Y}_{\epsilon_1,\epsilon_2}}{M-1}\,
\frac{n^{\rm X}_{\epsilon_1,\epsilon_2}
F^{\rm X}_{\epsilon_1,\epsilon_2}}{\Phi_{\epsilon_1,\epsilon_2}},
\end{equation}
and
\begin{equation}
\omega^{\rm XX}_{\epsilon_1,\epsilon_2}=
\frac{n^{\rm X}_{\epsilon_1,\epsilon_2}-1}{M-1}\,
\frac{n^{\rm X}_{\epsilon_1,\epsilon_2}
F^{\rm X}_{\epsilon_1,\epsilon_2}}{\Phi_{\epsilon_1,\epsilon_2}},
\end{equation}

\section{Stationary probabilities in the weak mutation limit}
\label{app:A}

\subsection{Unconditional imitation and proportional update}

According to the payoffs obtained in Appendix~\ref{app:B}:
\begin{enumerate}[(i)]
\item $P_D(m,0)>P_C(m,0)$ for all $0<m<M$, so D always invades C, but C
never invades D.
\item $P_C(m,M-m)>P_J(m,M-m)$ for all $0<m<M$, provided $r>1+(n-1)d$
(the rock-paper-scissors condition), so under this assumption C always
invades J, but J never invades C.
\item $P_J(0,j)>P_D(0,j)$ for all $0<j<M$, so J always invades D, but D
never invades J.
\end{enumerate}
Therefore
\begin{equation}
\mathbf{Q}=
\begin{pmatrix}
-1 & \phantom{-}0 & \phantom{-}1 \\
\phantom{-}1 & -1 & \phantom{-}0 \\
\phantom{-}0 & \phantom{-}1 & -1 
\end{pmatrix}.
\end{equation}
This implies
\begin{equation}
\alpha_{\rm C}= \alpha_{\rm D}= \alpha_{\rm J}=\frac{1}{3}.
\label{eq:ui1}
\end{equation}

On the other hand, if $r<1+(n-1)d$ neither C invades J nor vice-versa, so
in this case
\begin{equation}
\mathbf{Q}=
\begin{pmatrix}
-1 & \phantom{-}0 & 0 \\
\phantom{-}1 & -1 & 0 \\
\phantom{-}0 & \phantom{-}1 & 0 
\end{pmatrix},
\end{equation}
which implies
\begin{equation}
\alpha_{\rm C}= \alpha_{\rm D}=0,\quad \alpha_{\rm J}=1.
\label{eq:ui2}
\end{equation}

\subsection{Moran Process}
\label{sec:appMoran}

The Moran process for a population with two strategies defines a
birth-death process with two absorbing states. The details of the
calculation of $\rho_{\rm YX}$ can be found in \cite{hauert:2007}
and follow standard formulae for this kind of processes \cite{karlin:1975}.
Summarizing, if we denote $P_{\rm YX}(m)$ the payoff received by
a type Y individual when the population is made of $m$ Y individuals
and $M-m$ X individuals, then
\begin{equation}
\rho_{\rm YX}^{-1}=\sum\limits_{m=0}^{M-1}q_m, \qquad
\rho_{\rm XY}=q_{M-1}\rho_{\rm YX},
\end{equation}
where $q_0=1$ and
\begin{equation}
q_m =q_{m-1}\frac{1-s+sP_{\rm XY}(M-m)}{1-s+sP_{\rm YX}(m)}, \qquad
0<m<M.
\end{equation}
Payoffs $P_{\rm XY}(m)$ and $P_{\rm YX}(m)$ are obtained from the
formulae of Appendix~\ref{app:B}.
The maximum value of the selection strength $s$ is given by
\begin{equation}
s_{\rm max}=\frac{1}{1-\min\limits_{{\rm XY},m}P_{\rm XY}(m)}.
\end{equation}

\bibliography{evol-coop-fp}

\end{document}